\documentclass[]{jfm}

\usepackage{graphicx}
\usepackage{newtxtext}
\usepackage{newtxmath}
\usepackage{natbib}
\usepackage{hyperref}
\hypersetup{
    colorlinks = true,
    urlcolor   = blue,
    citecolor  = black,
}

\newcommand{\RomanNumeralCaps}[1]
\linenumbers

\title{Direct Numerical Simulations of Ice-Ocean Boundary Turbulence}

\author{Ken X. Zhao\aff{1}
  \corresp{\email{kenzhao@unc.edu}},
  Tomas Chor\aff{2}, 
 Eric Skyllingstad\aff{3}, Jonathan Nash\aff{3}, Madelaine Rosevear\aff{4}, \and Craig McConnochie\aff{5}}

\affiliation{\aff{1}Dept. of Earth Marine and Environmental Sciences, University of North Carolina at Chapel Hill, NC, USA
\aff{2}Dept.\ of Atmospheric and Oceanic Science, University of Maryland at College Park, College Park, MD, USA
\aff{3} College of Earth, Ocean, and Atmospheric Sciences, Oregon State University, Corvallis, OR, USA
\aff{4} Institute of Marine and Antarctic Studies, University of Tasmania, Hobart, Tasmania, AUS
\aff{5}Department of Civil and Environmental Engineering, University of Canterbury, Christchurch, NZ
}
\begin{document}
\maketitle

\begin{abstract}
Turbulent heat and freshwater transport at ice-ocean interfaces controls glacier and iceberg melt rates, yet the underlying physics remains poorly constrained. Parameterizations that assume shear boundary layer scaling are commonly used, which neglects meltwater buoyancy-driven convective processes. Using Direct Numerical Simulations with realistic salt diffusivity, which is critical for representing the thin solutal boundary layer ($\delta_S \approx 0.4$ mm) and resulting convective instabilities, we investigate ice-ocean boundary layer turbulence across varying temperature, salinity, stratification, external velocity, and interfacial slope angles. Our simulations agree with laboratory measurements of melt rate and interfacial temperature. In the absence of external flows, we find no transition from buoyancy-controlled to shear-controlled regimes and convection is important even at near-horizontal slopes. External shear becomes significant only when it is strong enough to thin the thermal and solutal boundary layers, which starts influence melting substantially above background flow speeds of 5 cm/s. Understanding how shear and convection compete to determine the ice-ocean diffusive boundary layer enables accurate melt rate predictions across the parameter space relevant to ice shelves and marine-terminating glaciers.
\end{abstract}

\section{Introduction}\label{sec:introduction}

Recent atmospheric and oceanic warming has led to the enhancement of thermal forcing and turbulence at the ice-seawater interfaces of tidewater glaciers and ice shelves \citep{Holland08, Straneo13, Wood18, Cowton18, Slater22}. At submerged ice-ocean interfaces, turbulent plumes arise from both subglacial hydrological outlets and ambient melt processes. Together, these determine the efficiency of heat and salt fluxes at the ice-ocean boundary layer and drive the entrainment and upwelling of deep, warmer and saltier waters towards the surface \citep{Straneo15, Jackson17, Jackson20, Slater22}. The vertical distribution of this freshwater flux sets near-surface water mass properties in polar oceans, which influence large-scale ocean overturning circulation \citep{Dukhovskoy19, Li23}.

Current climate and ocean models must parameterize these ice-ocean interactions due to the computational challenge of resolving turbulent boundary layer processes at glacial melt scales. The turbulent ice-ocean boundary layer is typically represented by plume theory for the outer layer and melt parameterizations for the inner boundary layer \citep{Mcphee08,Wells08, McConnochie15, Jenkins19,Malyarenko20}. These theories rely on plume-integrated conservation laws of momentum, mass, and buoyancy, combined with inner boundary layer budgets of heat and salt. While there is ongoing work on entrainment and boundary layer drag parameterizations \citep{Kimura16, Ezhova18,Jackson20,Zhao23GRL}, accurately estimating heat and salt fluxes under varying geometric and hydrodynamic forcing remains crucial for constraining melt rates.

However, recent observations reveal a fundamental failure of existing theory. 
Ambient melt rates at LeConte Glacier, Alaska reach 1--10 meters/day across the 
entire submarine terminus, which is an order of magnitude greater than shear-driven 
parameterizations predict in regions far from discharge plumes 
\citep{Jackson20, Sutherland19}. This discrepancy is most acute in quiescent 
conditions where background currents are weak: \citet{Weiss2025} and 
\citet{Nash2024} demonstrate that observations in strong-shear environments 
are broadly consistent with shear-based parameterizations, but that the same 
framework fails catastrophically in the absence of significant background flow \citep{Jackson20, Zhao2024}, 
suggesting the dominant physical mechanism driving melt under quiescent conditions 
remains unidentified. This undermines confidence in projections of glacier retreat 
and ocean circulation in polar regions, particularly since discharge plumes 
(which arise from meltwater exiting subglacial hydrological outlets) typically 
cover only a small fraction of total glacier area, and most parameterizations 
predict ambient melt rates outside of these plumes to be much lower than observed 
(often centimeters per day; \citealt{Fried15, Carroll16, Zhao21}). Both 
face-averaged and total observed melt rates therefore substantially exceed 
parameterized predictions wherever background currents are insufficient to 
sustain shear-driven melting.

Ambient melt plumes, driven by buoyant convection from melting and externally-sourced turbulence and mean circulation \citep{Straneo15, Jackson20}, are critical to near-glacier dynamics. Horizontal velocities can reach tens of cm/s and vary significantly across fjords and seasons, influenced by water mass transformation and complex interactions between ocean currents and evolving glacier morphology \citep{Sutherland14, Cowton15, Straneo15, Slater18, Jackson20, Zhao21, Zhao21c, Zhao22, Zhao23, Zhao23GRL}. In ice shelf cavities larger scale circulation can similarly arise from tidal currents, cavity circulation, and plume-driven convection \citep{Rosevear25}. Observational data in both Northern and Southern hemisphere polar regions supports a transfer function merging velocity-dependent (shear-dominated) and velocity-independent (buoyancy-dominated) melt regimes based on empirical fits. This yields significantly higher baseline buoyancy-dominated melt rates than previous literature (e.g., \cite{Kerr15}) but provides no physical explanation for these elevated rates.

Previous direct numerical simulations (DNS) and laboratory experiments \citep{Gayen16, Mondal19, McConnochie15, McConnochie17, McConnochie17b, Kerr15, Wells08} have provided valuable insights into ice-ocean boundary layer dynamics, but uncertainties remain regarding appropriate parameterization of the diffusive boundary layer thickness and resulting melt rates. The present study advances beyond \cite{Gayen16} and \cite{Mondal19} by incorporating realistic salt diffusivities (Schmidt number $\mathrm{Sc} = \nu/\kappa_S = 2500$, compared to $\mathrm{Sc} = 150$ used in \citealt{Gayen16} and \citealt{Mondal19}), which we find significantly influence boundary layer momentum, shear stresses, and melt rates.

In thermally-forced melt plumes where buoyancy is dominated by temperature anomalies, turbulent boundary layer transitions from convective to shear-driven regimes have been documented \citep{Ke23, Grossmann11}, with recent work further characterizing these transitions in related configurations \citep{Wells23, Howland23, Yang23, Couston24}. However, for the double-diffusive ice-ocean boundary layer with realistic haline diffusivities, the nature of such transitions remains unknown and is a central focus of this study.

This study aims to reconcile the existing mismatch between theories and observations \citep{Jackson20,Sutherland19} and answers the following questions.
    \begin{itemize}
        \item
How does turbulence at the edge of the diffusive boundary layer constrain the melt rate? 
\item How does the melt rate respond to the presence of external shear, convective turbulence, or both simultaneously? 
\item How can these effects be incorporated into an effective parameterization?
\end{itemize}

In this study, we propose a physically-motivated melt parameterization that accounts for both buoyancy-driven and shear-dominated melt regimes, aligning with existing theories, observations, and laboratory experiments.
This study presents direct numerical simulations (DNS) of turbulent convection and dissolution at a vertical ice face in contact with a uniform ocean. Building on \citealt{Gayen16} and \citealt{Mondal19}, we restrict the simulations to approximately laboratory scales to enable direct comparison with previous experiments and ensure all scales of motion are adequately resolved.
Our approach enhances resolution across the entire diffusive boundary layer, uses realistic salinity diffusivities, and incorporates the melt rate as an interfacial retreating wall boundary condition; we argue that these factors are essential for capturing the correct dynamics. Critically, understanding the character of turbulence (whether convectively or shear-driven) is essential for determining how the boundary layers are perturbed under natural geophysical conditions, 
and therefore for constructing a parameterization that remains valid across the 
full range of oceanic forcing. We explore a wide parameter space, including 
variations in salt diffusivity, thermal and salinity forcing, interface slope 
angle, external velocity, and stratification, and investigate the momentum and 
scalar balances. These insights inform an updated melt parameterization, grounded 
in accurate predictions of the diffusive buoyancy boundary layer, advancing our 
understanding of melt dynamics and ice--ocean interactions under geophysical 
conditions.

\section{Problem Formulation and Numerical Methods}\label{sec:numerics}

\subsection{Problem setup}

We solve the incompressible, nonhydrostatic Navier-Stokes equations using a Bousinessq approximation along with conservation of mass, heat, and salt, along with a linear equation of state.
The momentum equations are 

\begin{subequations}
\begin{align}
    \frac{\partial u}{\partial t} + (\vec{u}\cdot \vec{\nabla}) u  &= -\frac{1}{\rho_0} \frac{\partial p}{\partial x} + \nu \nabla^2 u + \frac{\Delta \rho}{\rho_0}  g \cos \theta \,, \\
    \frac{\partial v}{\partial t} + (\vec{u}\cdot \vec{\nabla}) v  &= -\frac{1}{\rho_0} \frac{\partial p}{\partial y} + \nu \nabla^2 v \,, \\
    \frac{\partial w}{\partial t} + (\vec{u}\cdot \vec{\nabla}) w  &= -\frac{1}{\rho_0} \frac{\partial p}{\partial z} + \nu \nabla^2 w - \frac{\Delta \rho}{\rho_0}  g \sin \theta \,, \\
    \nabla \cdot \vec{u} &= 0 \,, \\
    \frac{\partial T}{\partial t} + (\vec{u}\cdot \vec{\nabla}) T &= \kappa_T \nabla^2 T \,, \\
    \frac{\partial S}{\partial t} + (\vec{u}\cdot \vec{\nabla}) S &= \kappa_S \nabla^2 S \,, \\
    \frac{\Delta \rho }{\rho_0} &= -\alpha (T- T_0) + \beta (S-S_0).
\end{align}
\end{subequations}

For simplicity, the flow velocity $\vec{u} = (u,v,w)$ is defined in the interfacial reference frame, i.e., in the wall-normal ($x$), spanwise ($y$), and wall-tangent/vertical ($z$) directions, respectively, with $x=0$ located at the ice interface. $p$ is pressure, $T$ is temperature, $S$ is salinity, $\Delta \rho = \rho - \rho_0$ is the density anomaly relative to the fixed reference density $\rho_0$, $T_0$ is the reference temperature, $S_0$ is the reference salinity, $g$ = 9.81 m s$^{-2}$ is the gravitational acceleration, $\alpha = 3.87 \times 10^{-5}$ $^\circ \text{C}^{-1}$ and $\beta = 7.86 \times 10^{-4}$ $\text{psu}^{-1}$, are the coefficients of thermal expansion and haline contraction \citep{Jenkins11}. The slope angle $\theta$ is defined relative to the horizontal direction (vertical thus corresponds to $\theta=90^\circ$). We use realistic values for the molecular viscosity $\nu = 1.8 \times 10^{-6}\,\mathrm{m^2\,s^{-1}}$, the molecular diffusivity of heat $\kappa_T = 1.3 \times 10^{-7}\,\mathrm{m^2\,s^{-1}}$ (Prandtl number $\mathrm{Pr} = \nu / \kappa_T = 14$), and the molecular diffusivity of salt $\kappa_S = 7.2 \times 10^{-10} \, \mathrm{m^2 \, s^{-1}}$ (Schmidt number $\mathrm{Sc} = \nu / \kappa_S = 2500$), except in simulations specifically designed to test sensitivity to these parameter variations. We note that $\nu$ and $\kappa_T$ vary with temperature in 
reality; however, both are treated as constant throughout this study, evaluated 
at representative near-freezing conditions. Previous DNS studies of ice–ocean boundary layers have typically used reduced Schmidt numbers for computational tractability (e.g., $\mathrm{Sc}=500$ in \cite{Mondal19} and $\mathrm{Sc}=100$ in \cite{Gayen16}).The importance of using realistic values for $\kappa_S$ will be discussed in the next subsection.

The conservation laws of heat and salt at the ice interface ($x=0$) help us constrain the melt rate (see e.g., \citealt{Josberger81, Holland99, Wells08, Kerr15}). 
The boundary conditions at the ice can be expressed as
\begin{subequations}\label{meltrate_bc}
\begin{align}
T_i &=  \lambda_1 S_i + \lambda_2 + \lambda_3 z \approx \lambda_1 S_i , \label{meltrate_bc_a}\\
\frac{\rho_w}{\rho_i} u(x=0) \equiv \dot{m} 
&= - \rho_i^{-1} L^{-1} \rho_w c_w \kappa_T \frac{\partial T}{\partial x_\perp} \label{meltrate_bc_b}\\
&= -  \rho_i^{-1} S_i^{-1} \rho_w \kappa_S \frac{\partial S}{\partial x_\perp}. \label{meltrate_bc_c}
\end{align}
\end{subequations}

Here $T_i$ and $S_i$ are the temperature and salinity at the ice interface. 
The coefficients $\lambda_1 = -5.73 \times 10^{-2}$ $^\circ$C psu$^{-1}$, 
$\lambda_2 = 8.32 \times 10^{-2}$ $^\circ$C, and 
$\lambda_3 = 7.61 \times 10^{-4}$ $^\circ$C m$^{-1}$ 
are coefficients in a linearized expression for the freezing point of seawater as a function of salinity and pressure: specifically the freezing point slope, offset, and depth (pressure) dependence coefficient, respectively \citep{Jenkins11}.

This formulation assumes that the diffusive salinity flux and the salinity within the ice are negligible, and that the ice is at the freezing point (i.e., there is no sensible heat flux into the ice; see \citealt{Kerr15}). 
Here, $\rho_i$ and $\rho_w$ are the densities of ice and seawater, respectively. 
The specific heat capacity of seawater is 
$c_w = 3974~\mathrm{J\,kg^{-1}\,^\circ C}$,
and $L = 3.35 \times 10^5~\mathrm{J\,kg^{-1}}$ is the latent heat of fusion for ice. 
These empirical values are consistent with those used in previous studies for near-freezing temperature seawater environments
\citep{Jenkins10, Sciascia13, Cowton15}. 
Note that we have assumed a constant density ($\rho_w$) and specific heat capacity ($c_w$) for seawater, although both should vary with the salinity of the boundary layer seawater.

\subsection{Numerical Methods and Model Configuration}

We solve Eqs.\ (2.1)--(2.3) numerically using Oceananigans.jl,
a numerical fluid dynamics solver using graphics processing unit (GPU) acceleration \citep{Ramadhan20}.
The simulations in this paper use a finite-volume spatial discretization with a fifth order Weakly Essentially Non-Oscillatory (WENO) advection scheme and a second-order Adams-Bashforth time discretization.
The ice-ocean configuration used in these direct numerical simulations does not use a turbulent closure as it seeks to resolve all turbulent scales of motion.

Our simulations are conducted in a three-dimensional domain of size
$L_x \times L_y \times L_z$. The domain is periodic in the $y$-direction
and bounded by rigid, no-slip walls at the top and bottom, as well as at
the $x=0$ and $x=L_x$ boundaries.
The reference configuration uses
$L_x = 1.5~\mathrm{m}$, $L_y = 0.1~\mathrm{m}$, and $L_z = 3~\mathrm{m}$,
discretized on a uniform grid of $400 \times 100 \times 1500$ points.

To resolve the ice-ocean boundary layer while maintaining computational efficiency, we employ grid stretching in the wall-normal direction.
The $x$-grid uses hyperbolic tangent stretching to concentrate resolution near the ice boundary, achieving spacing of approximately $\Delta x \approx 10~\mathrm{\mu m}$ (for the first 100 gridpoints in $x$) near the interface, gradually expanding to $\Delta x \approx 1$ cm in the quiescent far field.
The resolution is found to be sufficient to resolve the Batchelor scale within the diffusive boundary layer and the Kolmogorov scale throughout our domain.

At the ice boundary ($x = 0$), we impose interactive boundary conditions for temperature, salinity, and wall-normal velocity (assuming a velocity of a retreating wall due to melt rate; this is discussed further in Section 3) based on Eqs.\ 2.2--2.3b. The boundaries at $x=L_x$, $z=0$ and $z=L_z$ employ zero-flux conditions for temperature and salinity. To prevent spurious reflections, we implement a sponge layer in the outer part of the domain (wall opposite the ice boundary). The sponge applies Rayleigh damping to the prognostic fields with spatially-varying damping coefficient $\lambda(x) = \lambda_0 \mathcal{M}(x)$, where $\lambda_0$ is the maximum damping rate and $\mathcal{M}(x)$ is a smooth masking function that transitions from zero to unity across the sponge layer extent $x_1 \leq x \leq x_0$.

The simulations reported in this study are summarized in Table~\ref{tab:param_runs}. All simulation code and configuration files are available in a permanent repository at \texttt{github.com/zhazorken/iceplume}.

\begin{table}
  \begin{center}
\def~{\hphantom{0}}
  \begin{tabular}{lcccccc}
      \hline
      Case & $T_\infty$ (deg.\ C) & $S_\infty$ (psu) & $\theta$ (deg.) & $v_\infty$ (cm~s$^{-1}$) & $N^2$ (s$^{-2}$) & $\dot{m}$ ($\mu$m~s$^{-1}$) \\[3pt]
      \hline
      Reference & 5.4 & 34.9 & 90 & ~0 & 0 & 3.9 \\[3pt]
      \multicolumn{7}{l}{\textit{Temperature variation (without moving boundary)}} \\
      T1 & 0.3 & 34.9 & 90 & ~0 & 0 & 0.8 \\
      T2 & 2.3 & 34.9 & 90 & ~0 & 0 & 1.9 \\
      T4 & 8.0 & 34.9 & 90 & ~0 & 0 & 5.3 \\[3pt]
      \multicolumn{7}{l}{\textit{Temperature variation (with moving boundary)}} \\
      T1m & 0.3 & 34.9 & 90 & ~0 & 0 & 0.9 \\
      T2m & 2.3 & 34.9 & 90 & ~0 & 0 & 2.3 \\
      T3m & 5.4 & 34.9 & 90 & ~0 & 0 & 4.5 \\
      T4m & 8.0 & 34.9 & 90 & ~0 & 0 & 8.0 \\[3pt]
      \multicolumn{7}{l}{\textit{Slope angle variation}} \\
      S1 & 5.4 & 34.9 & ~15 & ~0 & 0 & 3.2 \\
      S2 & 5.4 & 34.9 & ~30 & ~0 & 0 & 3.1 \\
      S3 & 5.4 & 34.9 & ~45 & ~0 & 0 & 3.2 \\
      S4 & 5.4 & 34.9 & ~60 & ~0 & 0 & 3.5 \\
      S5 & 5.4 & 34.9 & ~75 & ~0 & 0 & 3.6 \\
      S6 & 5.4 & 34.9 & 105 & ~0 & 0 & 4.1 \\
      S7 & 5.4 & 34.9 & 120 & ~0 & 0 & 4.5 \\[3pt]
      \multicolumn{7}{l}{\textit{External velocity variation}} \\
      V0 & 5.4 & 34.9 & 90 & ~~0 & 0 & 3.9 \\
      V1 & 5.4 & 34.9 & 90 & ~~5 & 0 & 4.3 \\
      V2 & 5.4 & 34.9 & 90 & ~10 & 0 & 4.9 \\
      V5 & 5.4 & 34.9 & 90 & ~25 & 0 & 8.1 \\[3pt]
      \multicolumn{7}{l}{\textit{Stratification variation}} \\
      N0 & 5.4 & 34.9 & 90 & ~0 & 0 & 3.9 \\
      N1 & 5.4 & 34.9 & 90 & ~0 & 5.0$\times 10^{-5}$ & 3.7 \\
      N2 & 5.4 & 34.9 & 90 & ~0 & 1.0$\times 10^{-4}$ & 3.6 \\
      N3 & 5.4 & 34.9 & 90 & ~0 & 2.0$\times 10^{-4}$ & 3.2 \\
      \hline
  \end{tabular}
  \caption{Summary of parameter variations explored in the sensitivity study. 
  The reference case corresponds to a vertical interface ($\theta = 90$ degrees) 
  with quiescent ambient conditions ($v_\infty = 0$, $N^2 = 0$) and 
  ambient temperature $T_\infty = 5.4$ deg.\ C at salinity $S_\infty = 34.9$ psu. 
  Each parameter group varies a single parameter while holding all others 
  at their reference values. Time-averaged melt rates $\dot{m}$ are computed 
  from the final 200s of each simulation after statistical steady state is achieved.}
  \label{tab:param_runs}
  \end{center}
\end{table}

\section{Results}\label{sec:results}
We now present a detailed analysis of the turbulent ice-ocean boundary layer structure and dynamics for the reference case (background temperature $T_\infty = 5.4$ deg.\ C, background salinity $S_\infty = 34.9$ psu, interface slope angle relative to horizontal $\theta = 90$ degrees, and background horizontal velocity $v_\infty = 0$), followed by systematic parameter variations. Section~\ref{sec:structures} examines the coherent turbulent structures that organize momentum and scalar transport near the interface. Section~\ref{sec:closure} develops a turbulence closure model that reconstructs temperature and salinity profiles across the diffusive and turbulent boundary layer regions. Section~\ref{sec:budgets} analyzes the momentum and scalar budgets to identify dominant physical processes and their spatial variation. Finally, Section~\ref{sec:sensitivity} explores parameter sensitivities across thermal forcing, interface geometry, external velocity, and stratification, providing the foundation for the unified melt rate parameterization developed in Section~\ref{sec:param}.

\subsection{Coherent Structures}
\label{sec:structures}

\begin{figure}
  \centerline{\includegraphics[width=0.94\linewidth,angle=0,trim={0pc, 0pc, 9pc, 0pc}, clip]{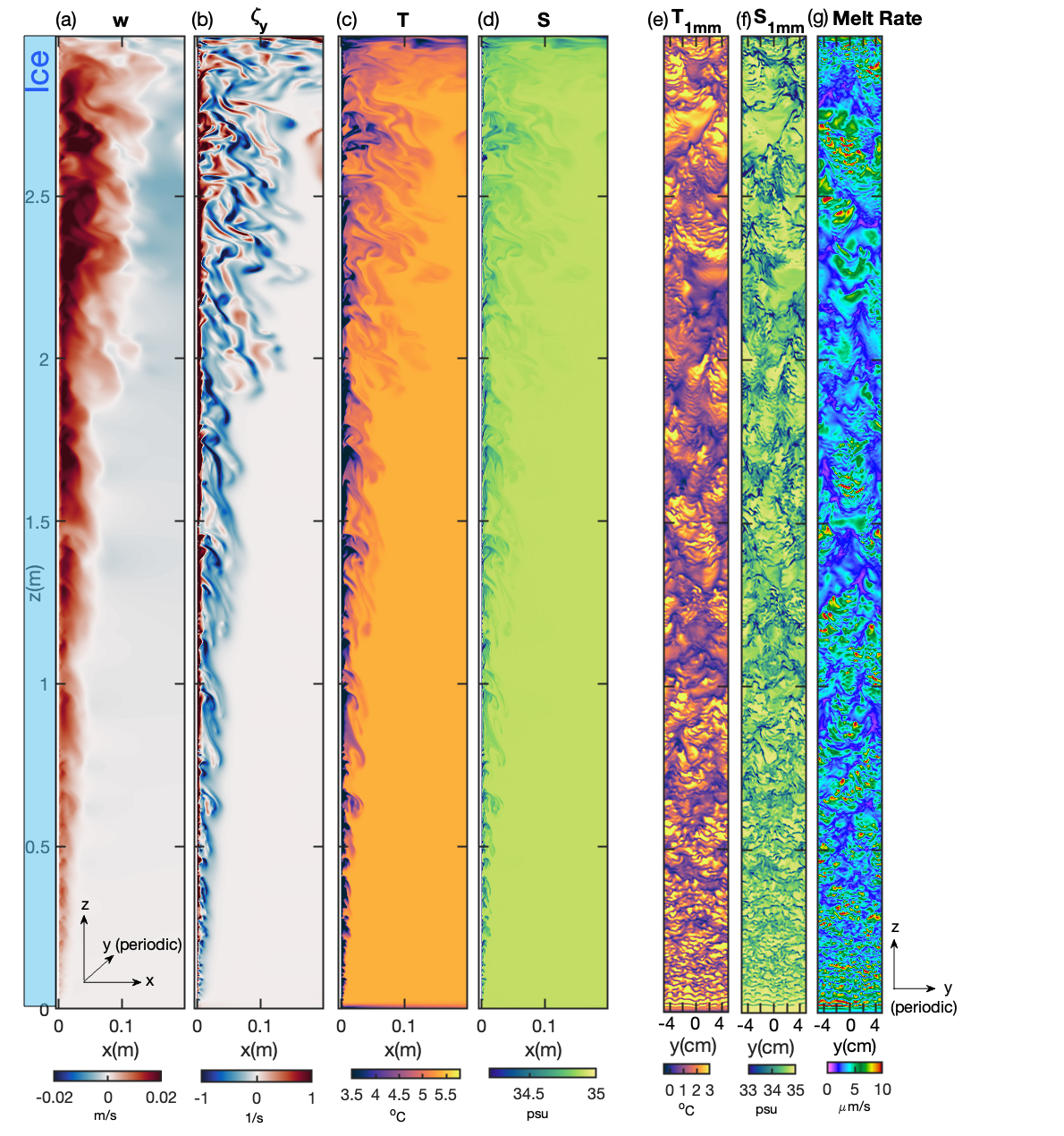}}
  \caption{Side view (x is normal to ice) of (a) w (vertical velocity), (b) $\zeta_y$ vorticity in y-plane (c) T (in situ temperature), (d) S (salinity) (e) temperature at yz-plane 1 mm away from the ice wall, (f) salinity at yz plane at 1 mm away from the ice wall, and (g) melt rate on the yz plane (um/s). All fields show the instantaneous variable at time t=600 seconds for reference case.}
\label{fig:1}
\end{figure}

\begin{figure}
\centerline{\includegraphics[width=1\linewidth,angle=0,trim={0pc, 0pc, 0pc, 0pc}, clip]{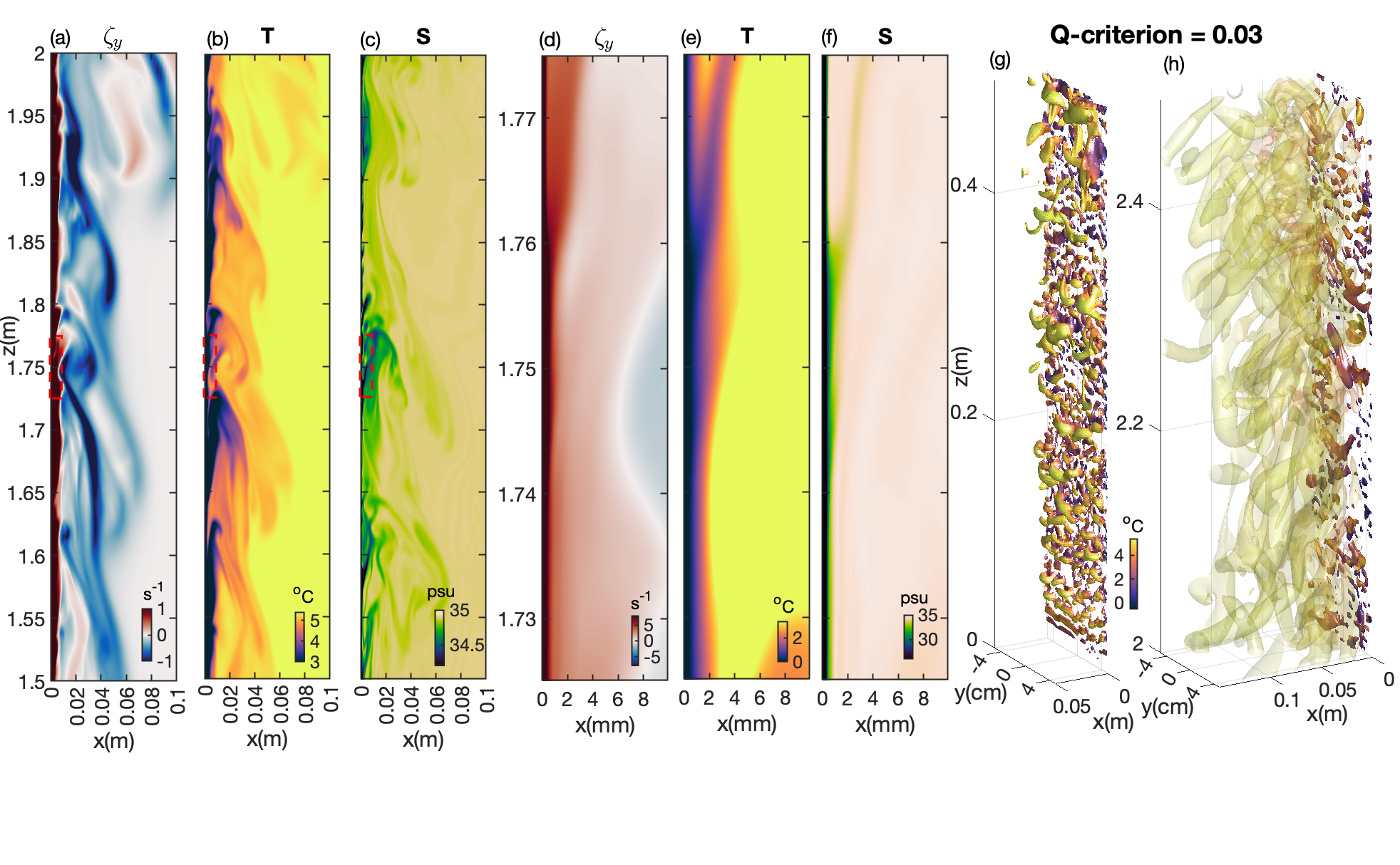}}
\caption{Instantaneous fields at $t=600$ s for the reference case in the $xz$-plane: 
  ($a$) vorticity in $y$-plane, 
  ($b$) temperature $T$, 
  ($c$) salinity $S$. 
  Panels ($d$)--($f$) show the same fields as ($a$)--($c$) zoomed to the millimeter scale near the ice interface. 
  Panels ($g$)--($h$) show three-dimensional vortex tube visualizations using isosurfaces of the 
  $Q$-criterion ($Q = 0.03$ s$^{-2}$), colored by vorticity magnitude.}
\label{fig:2}
\end{figure}

Figure~\ref{fig:1} provides an overview of the simulated boundary layer structure at $t = 600$ s. Side-view ($xz$-plane) snapshots show instantaneous fields of vertical velocity (panel a), spanwise vorticity (panel b), temperature (panel c), and salinity (panel d), revealing thin boundary layers adjacent to the ice interface ($\delta_T \approx 2$ mm, $\delta_S \approx 0.4$ mm; see Figure~\ref{fig:2}) and turbulent convective plumes. Cross-sectional views in the $yz$-plane at 1 mm from the ice wall (panels e-f) display the spanwise organization of the flow and scalar fields, while panel (g) shows the instantaneous melt rate distribution, which varies by a factor of 2-3 due to passing turbulent structures. The strong spatial variability in all fields highlights the inherently three-dimensional and time-dependent nature of the ice-ocean boundary.

The turbulent boundary layer exhibits organized coherent structures that control wall-normal momentum and scalar fluxes. Figure~\ref{fig:2}(a-c) shows instantaneous fields of vorticity, temperature, and salinity, revealing the complex interplay between convective plumes and background turbulence. Zoomed views near the interface (Figure~\ref{fig:2}d-f) highlight the thin boundary layers controlling scalar transfer. Three-dimensional visualizations of vorticity magnitude and $Q$-criterion 
isosurfaces (where $Q = \tfrac{1}{2}(|\boldsymbol{\Omega}|^2 - 
|\mathbf{S}|^2) > 0$, with $\boldsymbol{\Omega}$ and $\mathbf{S}$ the 
antisymmetric and symmetric parts of the velocity gradient tensor respectively, identifies regions where rotation dominates strain; 
\citealt{Hunt88}) in Figure~\ref{fig:2}g-h show vortex tubes and hairpin-like 
structures extending from the near-wall region into the outer flow. These 
coherent motions bear some resemblance to those in canonical wall-shear 
turbulence \citep{Robinson91, Adrian00} that arise from instabilities in the 
mean shear profile.

As hairpin vortices grow and lift from the interface, they generate characteristic sweep and ejection events: high-momentum fluid sweeps toward the wall ($u' < 0$, $w' > 0$), while low-momentum, buoyant fluid ejects outward ($u' > 0$, $w' < 0$), where primes denote fluctuations relative to the time-averaged mean (e.g., $u' = u - \overline{u}$). This asymmetric exchange produces Reynolds stress $(\overline{ u'w'})_x < 0$ that maintains the turbulent boundary layer against molecular diffusion away from the ice interface. Critically, these structures simultaneously influence momentum and scalar transport with varied effects. Sweep events bring warm, salty fluid toward the interface, thinning the viscous sublayer and enhancing instantaneous melt rate, but also mix away buoyant, cold meltwater, reducing the density anomaly driving convection. The melt rate thus emerges from the statistical balance between these competing processes, driven by vertical momentum and its variability as these sweeping eddies with intense positive vorticity anomalies near and within the diffusive boundary layer. These structures coincide with fine-scale boundary-layer gradients and enhanced variability in the scalar fields, particularly for salinity. Zoomed views of the near-wall region (Figure~\ref{fig:2}d--f) highlight the boundary layer structure for vertical velocity, temperature, and salinity.

\begin{figure}
  \centerline{\includegraphics[width=1\linewidth,angle=0,trim={0pc, 0pc, 0pc, 0pc}, clip]{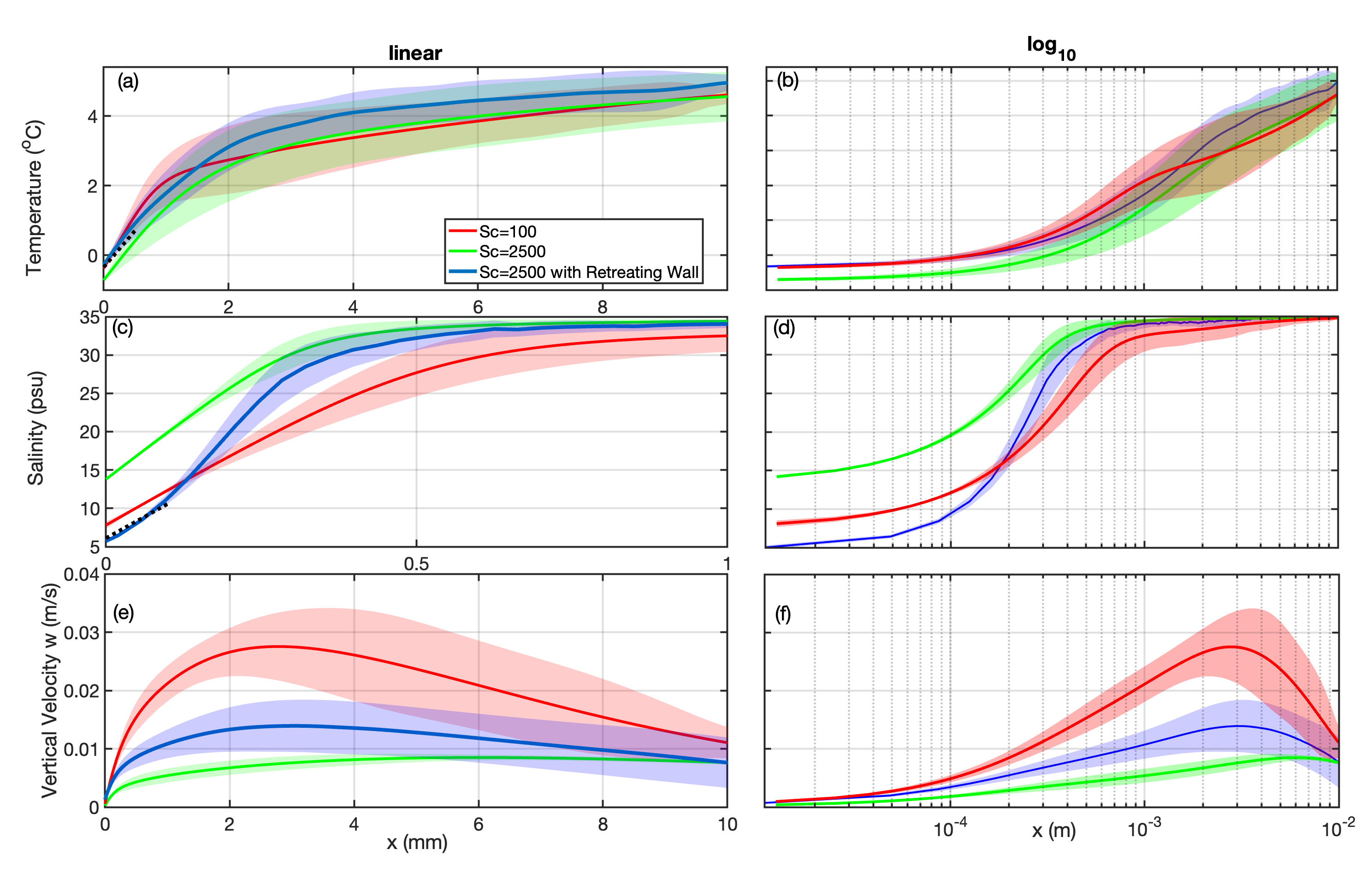}}
  \caption{Wall-normal ($x$) profiles of (a,b) temperature, (c,d) salinity, and (e,f) vertical velocity from simulations with varying Schmidt number and melt boundary conditions. Left column (a,c,e) shows linear scaling, while right column (b,d,f) shows logarithmic scaling to resolve near-interface structure. Three cases are compared: $Sc = 100$ (red), $Sc = 2500$ with a stationary boundary (green), and $Sc = 2500$ with retreating melt boundary (blue). Shading represents the root-mean square deviation of each parameter and these profiles are time- and vertically averaged.}
\label{fig:3}
\end{figure}

Figure~\ref{fig:3} presents wall-normal profiles of temperature, salinity, and vertical velocity, revealing the distinct length scales governing momentum and scalar transport near the ice interface. We define the three boundary layer thicknesses as follows: the thermal and solutal diffusive boundary layers ($\delta_T$ and $\delta_S$) are defined by the gradient method as $\delta_T = \Delta T ( \partial \overline{T} / \partial x |_{x=0} )^{-1}$ and $\delta_S = \Delta S ( \partial \overline{S} / \partial x |_{x=0} )^{-1}$, where $\Delta T$ and $\Delta S$ are the total thermal and solutal differences across the boundary layer. The viscous sublayer $\delta_\nu$ is defined as the wall-normal distance at which the magnitude of Reynolds stress $-(\overline{ u'w'})_x$ first exceeds the viscous stress $\nu \partial_{xx} \overline{w}$. By these definitions, the thermal diffusive boundary layer has thickness $\delta_T \approx 2.1$ mm, the solutal diffusive boundary layer $\delta_S \approx 0.4$ mm and the viscous sublayer $\delta_\nu \approx 4.7$ mm for the reference case.

These profiles demonstrate the critical role of Schmidt number and including the retreating (melting) wall velocity in controlling the salinity and momentum boundary layers, while the thermal boundary layer remains relatively insensitive to these variations. In particular, reducing the Schmidt number from $Sc = 2500$ to $Sc = 100$ substantially thickens the salinity boundary layer (panels c,d), as enhanced molecular diffusion of salt broadens the region of salinity adjustment. The retreating melt boundary induces a modest thinning of $\delta_S$ (approximately 10--15\%) and enhancement of interfacial fluxes (panel d), as the wall-normal advection $\dot{m} \partial_z S$ compresses the diffusive sublayer. By contrast, the thermal and momentum profiles show minimal sensitivity to boundary motion, as the advective contribution remains negligible compared to turbulent and molecular transport for these fields.

\begin{figure}
  \centerline{\includegraphics[width=0.8\linewidth,angle=0,trim={0pc, 0pc, 0pc, 0pc}, clip]{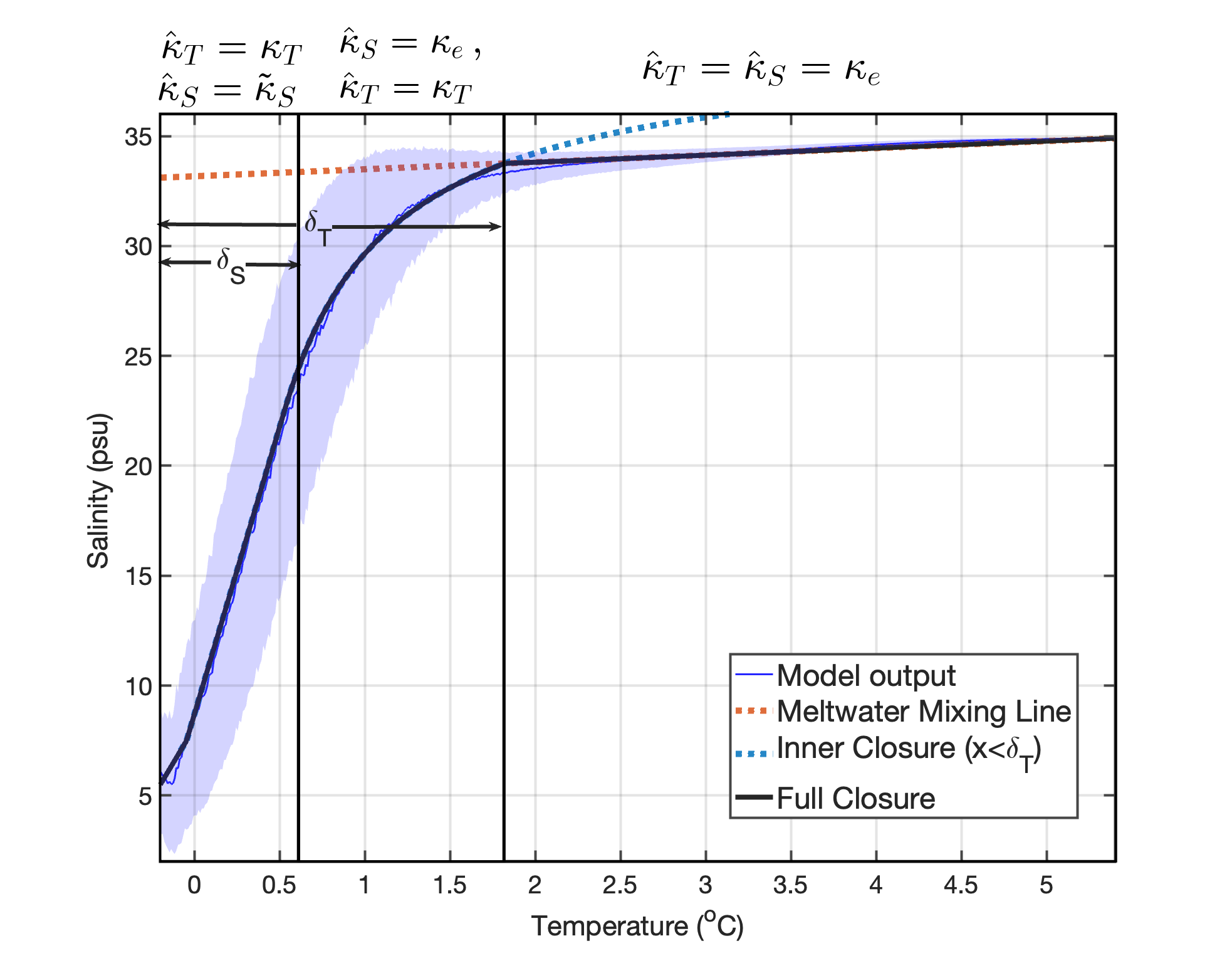}}
\caption{Temperature-Salinity diagram illustrating turbulent mixing and closure behavior near the ice-ocean interface. The diagram shows the evolution of water properties in T-S space, with salinity (psu) on the vertical axis and temperature ($^\circ$C) on the horizontal axis. The solid blue line represents the model output trajectory, with shaded regions indicating temporal variability. The dotted blue line denotes the inner closure approximation using Eq.\ \eqref{Eq33}. The red dashed line indicates the meltwater mixing line, connecting the ambient ocean properties to the interface conditions. The solid black line represents the full closure solution. Three distinct mixing regimes are identified in the panels above: (left) $\hat{\kappa}_T = \kappa_T$, $\hat{\kappa}_S = \kappa_S$, representing independent thermal and haline diffusion; (center) $\kappa_S = \kappa_e$, $\kappa_T = \kappa_T$, indicating enhanced salt transport; and (right) $\hat{\kappa}_T = \hat{\kappa}_S = \kappa_e$, representing uniform turbulent diffusion. The characteristic boundary layer thicknesses $\delta_T$ and $\delta_S$ are marked with vertical and horizontal bars, respectively, illustrating the different diffusively-dominated regions for heat and salt.}
\label{fig:4}
\end{figure}

\subsection{A Turbulence Closure for Predicting Boundary Layer Thicknesses}
\label{sec:closure}

The turbulent boundary layer structure described above can be distilled into 
three characteristic length scales ($\delta_S$, $\delta_T$, and $\delta_\nu$) whose relative magnitudes encode the competing roles of molecular diffusion, 
buoyancy, and momentum in controlling ice-ocean heat and salt transfer. The 
turbulence closure developed here exploits this hierarchy: because $\delta_S \ll 
\delta_T \leq \delta_\nu$, the salinity boundary layer sets the dominant resistance 
to buoyancy flux, and accurately predicting $\delta_S$ is sufficient to close the 
melt rate problem. This closure therefore forms the direct physical basis for the 
melt parameterization developed in Section~4.
Specifically, the following turbulence closure predicts temperature and salinity profiles by incorporating external temperature and salinity conditions, along with estimates of the freshwater diffusive layer thickness. While knowledge of the interface temperature $T_i$ and melt rate enhances the accuracy of the closure, we can also attempt to predict these quantities prognostically (see Section 3.5).

We first find the merged diffusivity to be the maximum of some eddy diffusivity function and the molecular diffusivity of temperature and salinity, respectively,
\begin{equation} \hat{\kappa}_S(x) = \max(\tilde{\kappa}_S, \kappa_e) , \quad \hat{\kappa}_T(x) = \max(\kappa_T, \kappa_e) . \label{eq:kappa_max} \end{equation} \label{closure1}

Within the thermal and salinity boundary layer, we find that the eddy diffusivity 
near the wall may be parameterized well using a quadratic scaling with distance 
from the wall, motivated by Prandtl's mixing length concept. Specifically, we 
adopt the functional form $\kappa_e \sim L^2 |dw/dx|$ with mixing length 
$L$ and assume the velocity gradient $dw/dx$ is approximately constant 
within the boundary layer, yielding:
\begin{equation} 
\kappa_e(x) = c_m \tilde{\kappa}_S \left(\frac{x}{\delta_S}\right)^2 \,, 
\label{eq:kappa_e} 
\end{equation} 
where $x$ is the distance from the wall, $\delta_S$ is the salinity boundary 
layer width, $\tilde{\kappa}_S$ is a characteristic diffusivity scale, and 
$c_m\approx 2.8$ is an empirical $\mathcal{O}(1)$ scaling coefficient. The 
parameterization is normalized by $\delta_S$ and $\tilde{\kappa}_S$ rather than 
their thermal counterparts because $\kappa_e$ is dynamically controlled by the 
salinity field: since $\delta_S \ll \delta_T$, the salinity boundary layer 
sets the scale of scalar mixing, and the curvature in the scalar profiles 
visible between $\delta_S$ and $\delta_T$ in Figure~\ref{fig:4} confirms that 
turbulent fluxes become significant only beyond $\delta_S$, leaving the 
innermost salinity sublayer diffusion-dominated. While Prandtl's original 
formulation was developed for momentum transport in wall-bounded shear flows, 
we find this quadratic scaling effectively captures the spatial variation of 
scalar eddy diffusivity within the high-Schmidt-number salinity boundary layer, 
where mixing occurs at scales much smaller than the momentum boundary layer 
(i.e., at the Batchelor scale).


While the regular molecular diffusivity $\tilde{\kappa}_S = \kappa_S$ is 
sufficient for capturing the salinity profile in Figure~\ref{fig:3}c in the 
absence of interface motion, the retreating melt boundary introduces a 
physically important modification that cannot be neglected. The wall-normal 
advection $\dot{m}\,\partial_x S$ associated with the melting interface actively 
redistributes salinity within the boundary layer, introducing curvature to the 
salinity profile (Figure~\ref{fig:3}) that is absent in the stationary-wall 
case. This curvature reflects a genuine dynamical effect: the meltwater flux 
compresses the inner salinity sublayer while simultaneously sustaining steeper 
outer gradients, effectively reducing the interfacial salinity $S_i$ and 
modifying the buoyancy anomaly that drives convection. Although the effect on 
the thermal and momentum budgets is negligible (owing to their much larger 
diffusivities), the salinity budget is acutely sensitive to this advective 
contribution, particularly at high melt rates where $\dot{m}\,\delta_S/\kappa_S 
\sim \mathcal{O}(1)$. Correctly capturing this effect is therefore essential for 
accurate melt rate predictions. For simplicity, we approximate the moving wall 
as having little effect on the inner part of the salinity boundary layer 
($x < \delta_S$) but reducing the effective salinity diffusivity in the outer 
part, and adopt the following modified salinity diffusivity to capture this 
influence:

\begin{equation}\label{Eq33}
\tilde{\kappa}_S \approx
\begin{cases}
            \kappa_S  \,, & \text{if } x < \delta_S  (\kappa_S/(2\dot{m}\delta_S)) \\
            \kappa_S (1+ \dot{m}\delta_S/\kappa_S)^{-1} \,, & \text{if } x \geq \delta_S  (\kappa_S/(2\dot{m}\delta_S)) \,.
        \end{cases}
\end{equation} 
Here, the proportion of the boundary layer affected and magnitude of effective 
reduction scales with the moving wall velocity (i.e., melt rate). This reduction 
captures the effect of wall-normal advection $\dot{m} \, \partial S/\partial x$ 
thickening the salinity boundary layer and sustaining steeper outer gradients 
(visible as increased curvature in Fig. 3c), which effectively decreases 
interfacial salinity $S_i$ compared to the stationary wall case. While $\dot{m}$ 
contributes negligibly to the temperature and momentum budgets, its role in 
redistributing buoyancy forcing away from the wall modifies the horizontal turbulent 
transport of salinity, which influences the vertical momentum budget as well.

To calculate the temperature and salinity profiles within the boundary layer, we start with the balance of fluxes at the interface. The gradients of temperature and salinity are expressed as:

\begin{subequations}
    \begin{equation}
        \frac{dT}{dx} = \frac{\rho_i L \dot{m}}{\rho_w c_w \hat{\kappa}_T(x)} \,, \label{eq:dTdx}
    \end{equation}
    \begin{equation}
        \frac{dS}{dx} = \frac{\rho_i S_i \dot{m}}{\rho_w \hat{\kappa}_S(x)} \,, \label{eq:dSdx}
    \end{equation}
\end{subequations}
where \( \rho_i \) is the density of solid ice, \( \rho_w \) is the density of seawater, \( L \) is the latent heat of fusion, \( S_i \) is the salinity at the interface, \( c_w \) is the specific heat capacity of seawater, and $\dot{m}$ is the melt rate (in m/s).

Using the closure from Eq.~\eqref{eq:kappa_max}, the temperature and salinity gradients can be integrated with respect to \( x \) to obtain the profiles for temperature and salinity. Outside the thermal boundary layer (where the turbulent diffusivity is larger than both molecular diffusivities of temperature and salinity), this can be merged with the meltwater mixing line, which assumes a linear relationship between temperature and salinity (and implicitly assumes $\kappa_e = \hat{\kappa}_S =  \hat{\kappa}_T$). The resulting merged profiles are defined as follows:

\begin{subequations}\label{fullclosure}
    \begin{equation}
        T(x) =
        \begin{cases}
           T_i + \int_{0}^{x} \dfrac{\rho_i L \dot{m}}{\rho_w c_w \hat{\kappa}_T(x')} \, dx' \,, & \text{if } x < \delta_T \\
           T_{\text{mix}}(x) \equiv \dfrac{T_\infty - \tilde{T}_i}{\delta_T} x + \tilde{T}_i \,, & \text{if } x \geq \delta_T
        \end{cases}
    \end{equation}
    
    \begin{equation}
        S(x) =
        \begin{cases}
            S_i + \int_{0}^{x} \dfrac{\rho_i S_i \dot{m}}{\rho_w \hat{\kappa}_S(x')} \, dx' \,, & \text{if } x < \delta_T \\
             S_{\text{mix}}(x) \equiv \dfrac{S_\infty - S_i}{\delta_S } x + S_i \,, & \text{if } x \geq \delta_T
        \end{cases}
    \end{equation}
\end{subequations}
where \(S_{\text{mix}}(x)\) and \(T_{\text{mix}}(x)\) represent the salinity and temperature profiles from the meltwater mixing line, respectively.

In Figure \ref{fig:4}, we show this closure compared to the simulation output $T$ and $S$ from the reference case (shading shows the root-mean square deviation from a mean temperature and salinity; the mean of salinity was calculated for each 0.01 degree bin of temperature). The meltwater mixing line is shown with a dotted red line. Here, we use a virtual interface temperature \(\tilde{T}_i = L/c_w + T_i = -84.6\) deg.\ C and \(S_i = 6.1\) psu. This merged closure from Eq.\ \eqref{fullclosure}
is shown in the solid black line. Note that this also allows us to predict the thermal and salinity boundary layer thickness by finding $\tilde{\kappa}_S = \kappa_e, \kappa_e = \kappa_T$.

\subsection{Momentum and Scalar Balance}
\label{sec:budgets}


\begin{figure}
  \centerline{\includegraphics[width=1\linewidth,angle=0,trim={0pc, 30pc, 0pc, 0pc}, clip]{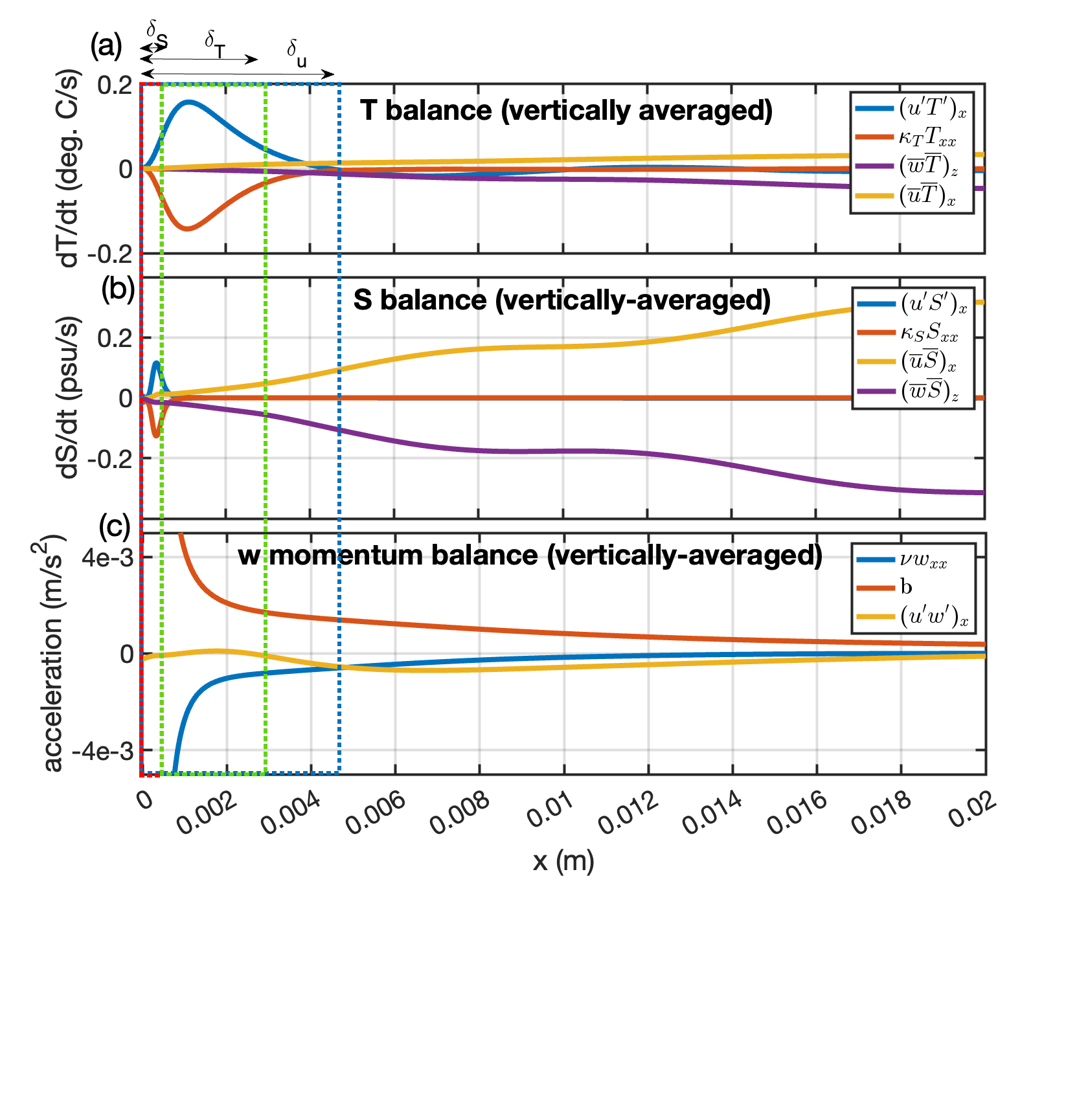}}
 \caption{Vertically and time-averaged budget terms for (a) temperature, (b) salinity, and (c) vertical momentum. Vertical dashed lines indicate the characteristic boundary 
layer thicknesses $\delta_T$, $\delta_S$, and $\delta_u$. 
Red lines denote the molecular diffusion terms for temperature and salinity 
($\kappa_T \partial_{xx} T$, $\kappa_S \partial_{xx} S$) and the buoyancy term, $b = g(\alpha T - \beta S)$. 
Yellow and purple curves denote the horizontal and vertical advection terms in the scalar budgets and yellow denotes the Reynolds 
stress term in the momentum budget. $\overline{\cdot}$ denotes a time average.}
\label{fig:5}
\end{figure}

\begin{figure}
\centerline{\includegraphics[width=1\linewidth,angle=0,trim={0pc, 0pc, 0pc, 0pc}, clip]{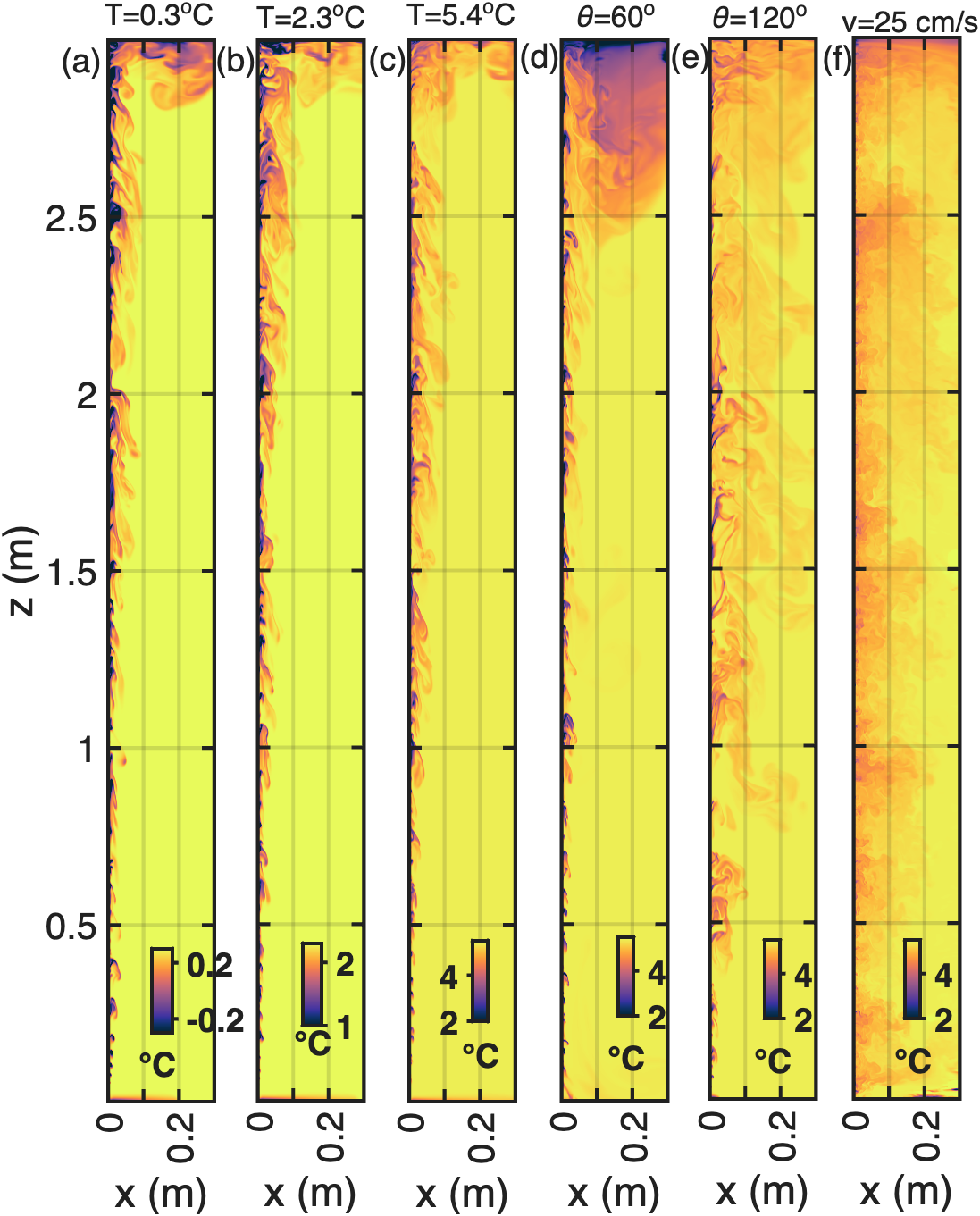}}
  \caption{Mid-y snapshot temperature sections $T(x,z)$ for varying thermal forcing and geometric configurations. Panels show cases with far-field temperature (a) $T_\infty = 0.3\,^\circ$C  (b) $T_\infty = 2.3\,^\circ$C, (c) $T_\infty = 5.4\,^\circ$C. Geometric variations include interface slope angles from (d) undercut ($\theta = 60^\circ$) to (e) overcut ($\theta = 120^\circ$) configurations. (f) external velocity $v_\infty = 25$ cm/s.}
\label{fig:6}
\end{figure}

\begin{figure}
  \centerline{\includegraphics[width=1\linewidth,angle=0,trim={0pc, 40pc, 0pc, 0pc}, clip]{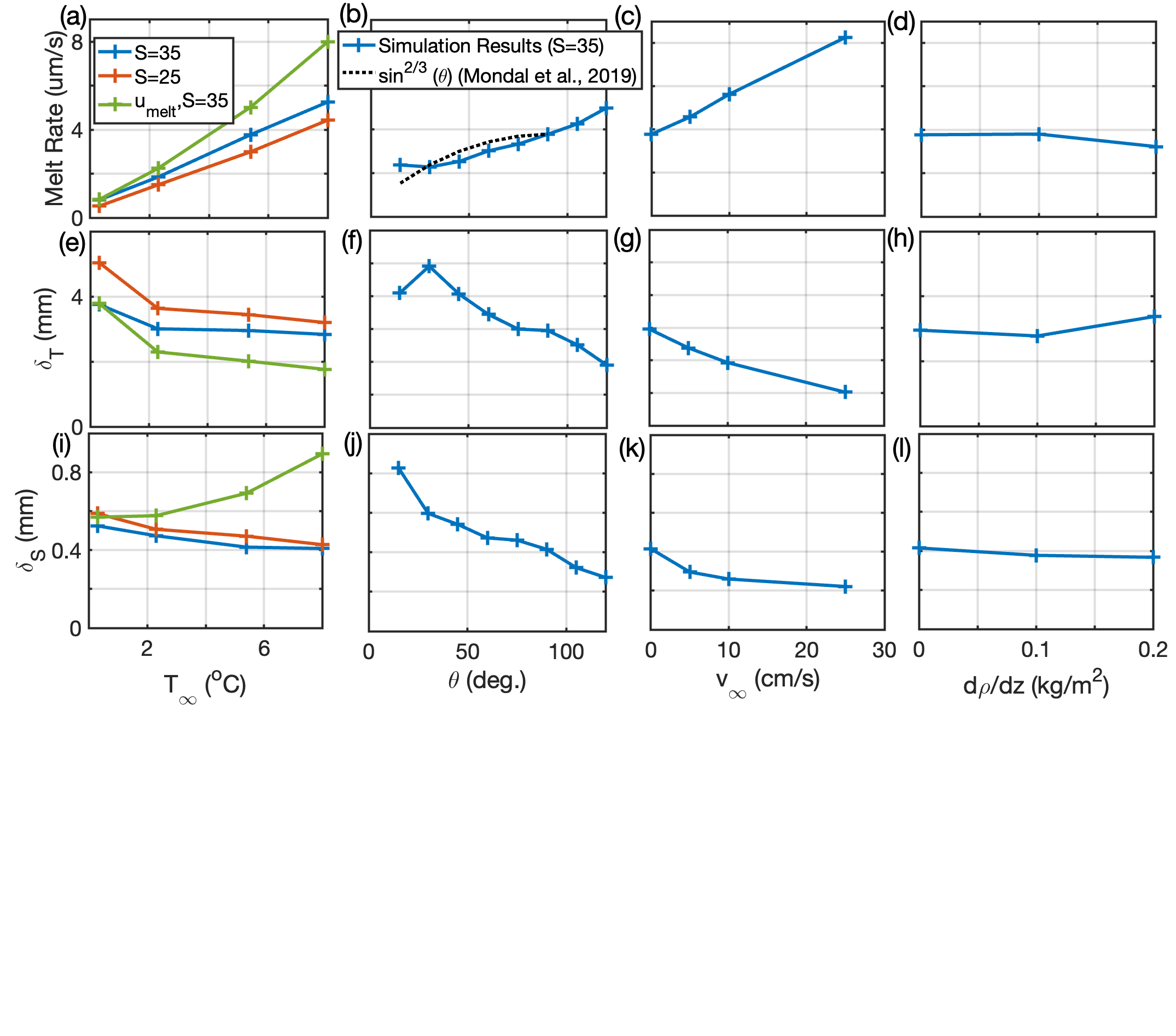}}
  \caption{Variation of melt rate (top row), thermal boundary layer thickness $\delta_T$ (second row), and solutal boundary layer thickness $\delta_S$ (third row) across parameter sweeps. The leftmost column shows sensitivity to salinity ($S = 25,35$ psu) and moving melt boundary ($u_{\text{melt}}$). The second column compares simulation results (blue) with the theoretical scaling $\sin^{2/3}\theta$ from \citet{Mondal19} (dashed black) as a function of interface slope angle $\theta$. The third and fourth columns show variations with external velocity $v_\infty$ and stratification $\partial_z{\rho}=-(\rho_0 g^{-1}) N^2$, respectively.} 
\label{fig:7}
\end{figure}

In this subsection, we quantify the wall-normal scalar and momentum balances. 
Figure~\ref{fig:5} shows the momentum and scalar balances for the reference case, highlighting the distinct structure and scales of the temperature, salinity, and momentum boundary layers, along with the dominant terms in each balance (Eqs.\ \eqref{eq:momentum_balance}--\eqref{eq:S_balance}). 
Beyond the viscous sublayer, the vertical momentum balance requires the divergence 
of the Reynolds stress to balance buoyancy forcing,
\begin{equation}\label{eq:momentum_balance}
    \partial_x \big( \overline{u'w'} \big) \sim g \beta \Delta S,
\end{equation}
where $\overline{\cdot}$ denotes a time average and $\Delta S$ represents the bulk salinity anomaly driving convection. Note that the haline contribution to buoyancy dominates over the thermal contribution ($\alpha \Delta T \ll \beta \Delta S$) throughout the boundary layer, justifying the use of salinity anomaly as the primary buoyancy driver.

Similarly, the scalar transport equations balance mean advection with turbulent 
flux divergence,
\begin{align}
    \overline{u}\,\partial_x \overline{T}
    + \partial_x \overline{u'T'}
    + \overline{w}\,\partial_z \overline{T}
    + \partial_z \overline{w'T'} &= 0, \label{eq:T_balance} \\
    \overline{u}\,\partial_x \overline{S}
    + \partial_x \overline{u'S'}
    + \overline{w}\,\partial_z \overline{S}
    + \partial_z \overline{w'S'} &= 0. \label{eq:S_balance}
\end{align}

Within the outer boundary layer, the turbulent fluxes 
$\overline{w'T'}$ and $\overline{w'S'}$ dominate vertical transport. 
Closer to the interface, however, molecular diffusion becomes dominant at 
different distances for temperature ($\delta_T \approx 2.1~\mathrm{mm}$) 
and salinity ($\delta_S \approx 0.4~\mathrm{mm}$), respectively.

The temperature and salinity balances both exhibit thin diffusive boundary layers near the wall where molecular diffusion dominates, transitioning to mean advection control in the outer region. Their structure is fundamentally similar, differing primarily in scale: the salinity boundary layer is much thinner than the thermal boundary layer due to the much lower molecular diffusivity of salt ($Sc = \nu/\kappa_S \approx 2500$ for seawater), a direct consequence of the large Lewis number ($Le = \kappa_T/\kappa_S \approx 180$). Within the thin salinity diffusive sublayer, molecular diffusion $\kappa_S S_{xx}$ dominates, while vertical advection of salinity becomes the dominant balancing term at the outer edge. This stark difference in diffusive length scales between heat and salt is fundamental to double-diffusive convection at ice-ocean interfaces.

The momentum balance reveals buoyancy forcing $b = g(\alpha T - \beta S)$ as the primary driver of near-wall flow within the buoyancy-affected region. Farther from the interface ($z > \delta_u$), Reynolds stress divergence $-\partial_x(\overline{u'w'})$ increases such that it is the primary term balancing buoyant production outside of the momentum boundary layer (instead of viscous dissipation). The momentum boundary layer thickness $\delta_u \approx 4.7~\text{mm}$ is significantly larger than $\delta_T$ and $\delta_S$.

A critical feature of these boundary layers is the proximity of buoyancy production of turbulent kinetic energy (TKE) to the interface. The density anomaly generated by differential diffusion occurs within a millimeter-scale region, resulting in substantially elevated frictional stress at the wall compared to canonical shear-driven boundary layers. Following the analysis of \citet{Gayen16}, we characterize this enhanced near-wall stress through an effective drag coefficient $C_d$, defined as $C_d = \tau_w / (\rho_0 U^2)$ where $\tau_w$ is the wall stress and $U$ is a characteristic velocity scale. Importantly, this drag coefficient differs from the bulk drag coefficient used in the three-equation melt model, where typical values are $\sim 2.5 \times 10^{-3}$; here the convectively-driven boundary layer generates substantially larger values. We find that $C_d$ varies in the range $0.1$--$2$ depending on the thickness and magnitude of the buoyancy anomaly (not shown), which itself is sensitive to the ambient temperature and salinity conditions. The extreme variability arises primarily from the salinity boundary layer dynamics: in the purely convective regime, the thin haline diffusive sublayer generates exceptionally strong density gradients and correspondingly large interfacial stresses. This elevated and highly variable drag has important implications for both the dynamics and for parameterizing ice-ocean momentum transfer in large-scale models, where constant drag coefficient assumptions may introduce significant errors.

The three boundary layer thicknesses ($\delta_T$, $\delta_S$, $\delta_u$) provide complementary proxies and diagnostics of the heat, salt, and momentum fluxes at the ice-ocean boundary. The thermal and salinity boundary layer thickness are the most directly relevant for melt rate predictions, as they together set the diffusive resistance (how strongly molecular diffusion limits 
the flux of heat and salt across the boundary layer) controlling buoyancy flux away from the wall and heat flux towards to the interface. The salinity boundary layer thickness $\delta_S$ governs the production of density anomalies and thus the strength of convective driving; its thinness relative to $\delta_T$ and $\delta_u$ is responsible for the elevated interfacial stresses observed in these simulations.

\subsection{Parameter Sensitivities}
\label{sec:sensitivity}

We investigate how the turbulent ice--ocean boundary layer responds to key physical and geometric controls. The parameter space explored in this study is summarized in Table~\ref{tab:param_runs}. In the first subsection, we first establish the importance of two foundational modeling requirements for an accurate representation of the physical system: the use of a realistic Schmidt number, $Sc = \nu/\kappa_S$, and the inclusion of melt-induced interface motion through a moving-boundary formulation. As we show below, both are essential for capturing the correct near-interface scalar structure and buoyancy forcing.

We then examine the sensitivity of the flow dynamics to: (i) interface geometry, including slope angles from undercut ($\alpha < 90^\circ$) to vertical ($\alpha = 90^\circ$) to overcut ($\alpha > 90^\circ$); (ii) far-field temperature $T_\infty$ and salinity $S_\infty$, which set the thermal and haline driving; (iii) external horizontal velocity $v_\infty$, representing ambient ocean currents; and (iv) background stratification. For each configuration, we quantify changes in the momentum and scalar budgets, turbulent fluxes, and melt rates relative to a reference case ($T_\infty = 5.4\,^\circ$C, $S_\infty = 34.9$ psu, $\alpha = 90^\circ$, $v_\infty = 0$, $N^2 = 0$).

Figure~\ref{fig:6} illustrates how selected parameter variations modify the buoyant melt plume, visualized using the temperature field $T(x,z)$. The six representative cases span the explored parameter space, contrasting weak thermal forcing ($T_\infty = 0.3\,^\circ$C; left column) with strong thermal forcing ($T_\infty = 2.3\,^\circ$C; right column), interface slope from undercut ($\alpha = 60^\circ$) to overcut ($\alpha = 120^\circ$), and a case with strong external velocity forcing ($v_\infty = 25$ cm s$^{-1}$) to highlight the transition from buoyancy-dominated to shear-influenced regimes.

\subsubsection{Realistic Schmidt Number and Moving Melt Boundary}

The Schmidt number $Sc = \nu/\kappa_S$ sets the relative diffusivities of 
momentum and salinity and therefore controls the thickness of the solutal 
boundary layer and the magnitude of near-wall density gradients. We examine 
$Sc \in \{100,\,500,\,2500\}$ to quantify how molecular transport properties 
influence scalar structure and turbulent mixing near the interface. Lower Schmidt 
numbers produce thicker salinity boundary layers, which reduce near-wall density 
gradients and weaken convective driving. Critically, at realistic oceanic values 
($Sc \approx 2500$), the salinity boundary layer becomes extremely thin 
($\delta_S \approx 0.38$ mm), confining strong density anomalies to a narrow 
region immediately adjacent to the interface and generating substantially more 
intense convective forcing than occurs at the reduced Schmidt numbers used in 
prior DNS. Therefore, realistic $Sc$ is a physical necessity rather for capturing the boundary layer dynamics accurately.

Figure~\ref{fig:3} demonstrates the strong sensitivity of the salinity boundary layer thickness to the Schmidt number. As $Sc$ decreases from 2500 to 100, $\delta_S$ increases markedly due to enhanced molecular diffusion of salt. This thickening weakens the near-wall density gradient, reduces buoyancy production, and lowers interfacial stress. The melt rate responds accordingly, decreasing at lower $Sc$ as convective driving is diminished. In contrast, the thermal boundary layer thickness $\delta_T$ remains relatively uniform across all cases, reflecting the comparitively larger thermal diffusivity. For the remainder of our simulations, we use the realistic value of $Sc$ and salinity diffusivity.

In addition, we explicitly account for the velocity of the melting interface through a moving-boundary formulation (in Eq.\ 2.3). This allows the meltwater flux and near-wall scalar gradients to remain dynamically coupled, rather than being artificially imposed at a fixed wall. Neglecting this kinematic boundary condition leads to systematic underestimation of buoyancy production and melt rates, particularly at high $Sc$.
Figure~\ref{fig:7} highlights the sensitivity of the system to the inclusion of a moving melt boundary. The melt boundary recession velocity $u_\mathrm{melt} \equiv u (x=0) \equiv \rho_i \rho_w^{-1}  \dot{m}$ introduces an additional advective contribution to wall-normal transport. We compare stationary and moving boundary conditions (in Figure~\ref{fig:7}a,e,i, labeled $u_{\text{melt}}$), and examine how interface motion modifies the near-wall velocity gradients and scalar fluxes.

The salinity balance near the interface is particularly sensitive to the advective term $u_\mathrm{melt} \partial_z S$, whereas the corresponding contribution is negligible in the thermal and momentum budgets because of their much larger diffusivities. The scalar budget reveals that the retreat of the melting boundary thickens the diffusive sublayer by up to 60\% (in the highest melt rate cases), which has an overall effect of decreasing the melt rate but increasing the vertical momentum. This effect on the solutal boundary layer and boundary layer dynamics for the range of ambient temperatures simulated in our study is therefore essential to capture for accurate representations of both melt rates and near-interface boundary layer dynamics.

\subsubsection{Varying Temperature and Salinity}

Ambient temperature $T_\infty$ directly controls the thermal driving and, in turn, the melt rate. As shown in Table~\ref{tab:param_runs} and Figure~\ref{fig:7}a, increasing $T_\infty$ from $0.3\,^\circ$C to $8.0\,^\circ$C monotonically increases the melt rate from $0.8~\mu$m~s$^{-1}$ to $8.0~\mu$m~s$^{-1}$ for the moving (melt) boundary cases (and up to $5.3~\mu$m~s$^{-1}$ for the stationary-boundary cases). The scaling is approximately proportional to the thermal forcing $\Delta T \equiv T_\infty - T_f(S_\infty)$, where $T_f(S_\infty)$ is the salinity-dependent freezing temperature. For the moving-boundary cases, the melt rate dependence is consistent with a $(\Delta T)^{4/3}$ scaling, in agreement with \citet{Mondal19} and Kerr \& McConnochie (2015), reflecting the classical turbulent natural convection scaling for the boundary layer thickness.

Figure~\ref{fig:7}a also shows that the melt rate is systematically lower at reduced ambient salinity ($S_\infty = 25$ psu) compared to the reference case ($S_\infty = 34.9$ psu), even for the same thermal forcing. This decrease arises from the proportionally smaller buoyancy flux, which weakens convective transport in the boundary layer.

The response of the scalar boundary layers is summarized in Figures~\ref{fig:7}e and \ref{fig:7}i. With increasing thermal forcing, the thermal boundary layer thins (Figure~\ref{fig:7}e), consistent with enhanced buoyancy-driven turbulence and stronger vertical heat transport. In contrast, the salinity boundary layer thickens (Figure~\ref{fig:7}i) as melt rates increase. This behavior is primarily a consequence of the retreating interface: as the wall recedes more rapidly, the advective contribution from the moving boundary becomes stronger, effectively broadening the region over which salinity is redistributed.

Together, these trends demonstrate that temperature and salinity influence the melt not only through their effect on the bulk buoyancy forcing, but also by reshaping the near-wall scalar boundary layers that regulate turbulent transport.

\subsubsection{Interface Slope Angle}

The ice-ocean interface inclination angle $\theta$ modifies the component of buoyancy forcing parallel and perpendicular to the interface. Melt rates (Table~\ref{tab:param_runs} and figure \ref{fig:7}b) exhibit a nonlinear dependence on slope angle (cases S1--S7 in Table~\ref{tab:param_runs}). For shallow angles ($\theta = 15$--$30^\circ$), melt rates remain relatively constant at $\dot{m} \approx 3.1$--$3.2~\mu$m~s$^{-1}$, with gentler interface slope angles requiring higher resolution to resolve the turbulent boundary layer. The melt rate gradually increases through intermediate angles ($\theta = 45$--$75^\circ$, $\dot{m} = 3.2$--$3.6~\mu$m~s$^{-1}$), reaching $\dot{m} = 3.9~\mu$m~s$^{-1}$ for the vertical reference case ($\theta = 90^\circ$), and continues to increase for overcut configurations ($\theta = 105$--$120^\circ$, $\dot{m} = 4.1$--$4.5~\mu$m~s$^{-1}$).

For overcut slope angles ($\theta > 90^\circ$), the orientation of the buoyancy forcing relative to the interface fundamentally alters the convective dynamics. In this configuration, the boundary layer becomes convectively unstable, leading to progressively higher melt rates as the slope angle increases. Notably, the approximately linear increase in melt with slope represents a clear departure from existing slope-dependent parameterizations (e.g., \citealt{Mondal19}).

Scalar transport exhibits expected corresponding variations. The thermal boundary layer thickness varies as $\delta_T \sim 1.5$--$2.5$ mm across the explored range, with thinner boundary layers at steeper angles and the thinnest boundary layers for overcut slopes. The solutal boundary layer thickness ranges from $\delta_S \approx 0.55$ mm at shallow angles to $\delta_S \approx 0.39$ mm at the steepest overhanging angles.

\subsubsection{External Horizontal Velocity}
An imposed ambient current $v_\infty$ (cases V0--V5) represents background ocean circulation parallel to the ice-ocean interface. Table~\ref{tab:param_runs} and figure \ref{fig:7}c demonstrate that melt rates increase approximately linearly with external velocity starting from the baseline convective melt rate: $\dot{m}$ rises from $3.9~\mu$m~s$^{-1}$ at $v_\infty = 0$ to $8.1~\mu$m~s$^{-1}$ at $v_\infty = 25$ cm~s$^{-1}$. This linear dependence is consistent with the most commonly-used parameterization of ice melting \citep{Jenkins19}, with the important distinction that convection sustains a non-zero baseline melt rate in the limit $v_\infty \to 0$.

The physical mechanism underlying this enhancement involves external velocity shearing away the thermal and solutal boundary layers. Increasing $v_\infty$ generates wall-parallel shear that intensifies turbulent mixing, reducing $\delta_T$ and $\delta_S$. Thinner boundary layers increase the temperature and salinity gradients at the interface, thereby enhancing heat and salt fluxes.

The momentum budget transitions from buoyancy-dominated ($v_\infty = 0$) to shear-influenced regimes as external velocity increases. 
For $v_\infty > 5$ cm s$^{-1}$, the magnitude of Reynolds stress magnitude exceeds the viscous shear within $\delta_T$ (not shown). 
Scalar boundary layer thicknesses decrease systematically with $v_\infty$ (Figure~\ref{fig:7}g,k). The thermal boundary layer thins from $\delta_T \approx 2.1$ mm at quiescent conditions to $\delta_T \approx 1.2$ mm at $v_\infty = 25$ cm~s$^{-1}$, and the solutal boundary layer exhibits proportionally similar thinning, from $\delta_S \approx 0.45$ mm to $\delta_S \approx 0.22$ mm. The linear dependence of melt rate on $v_\infty$ is broadly consistent with the parameterisation of \citet{Jenkins91}, though a key distinction is that the Jenkins scheme predicts zero melt at zero velocity, whereas the present results show a finite buoyancy-driven melt rate persisting in the quiescent limit.

\subsubsection{Stratification}



Ambient stratification $\partial_z{\rho}=-(\rho_0 g^{-1}) N^2$ (cases N0--N3) represents the background ocean density gradient. The stratification range explored here corresponds to moderate oceanic conditions relevant to glacier fjord environments \citep{Zhao2024, Nash2024, Weiss2025}, chosen to remain within the regime where buoyancy-driven boundary layer dynamics are not overwhelmed by ambient stratification. While strong stratification is known to strongly suppress melt rates (\citet{McConnochie2016} observed an ${\sim}80\%$ reduction at $N^2 = 0.08$~s$^{-2}$) the weaker stratification values tested here are physically motivated by observed glacier fjord conditions, and are too weak to trigger double-diffusive layering. 

Table~\ref{tab:param_runs} and figure~\ref{fig:7}d show that greater stratification weakly suppresses melt rates: increasing $\partial_z{\rho}$ from $0$ to $0.2$~kg~m$^{-3}$ reduces $\dot{m}$ from $3.9~\mu$m~s$^{-1}$ to $3.2~\mu$m~s$^{-1}$, an 18\% decrease. This weak sensitivity arises because the primary momentum and scalar balances occur within the millimeter-scale boundary layers, where background stratification has limited influence. The reduction in melt rate is attributable to enhanced entrainment of ambient fluid into rising meltwater plumes: stratification inhibits vertical motion, causing plumes to spread laterally and entrain more surrounding water, diluting the buoyancy anomaly and weakening the near-wall convective circulation. However, because the dominant resistance to heat transfer resides in the thin diffusive sublayer where stratification is negligible, the overall impact on melt rates remains modest.

\section{An Updated Ice Melt Parameterization}
\label{sec:param}

The preceding parameter sensitivity experiments reveal systematic dependencies of melt rate on thermal forcing, external velocity, stratification, and interface geometry. We now synthesize these observations into a unified parameterization by recognizing that all variations fundamentally influence through two physical quantities that capture most of the variability of melt rate over the relevant parameter regimes: the density difference driving the buoyancy, and the density boundary layer thickness controlling the rate of heat and salt diffusion at the boundary. As shown in Section~\ref{sec:budgets}, the haline contribution to density dominates within the diffusive boundary layer ($\alpha \Delta T \ll \beta \Delta S$), so the density boundary layer is effectively equivalent to the solutal boundary layer $\delta_S$ (confirmed explicitly in Figure~\ref{fig:8}a). The competition between convective and shear-driven processes in determining this boundary layer thickness provides the key to understanding the diverse parameter sensitivities.

Our simple proposed melt rate theory is controlled by the buoyancy flux at the ice-ocean interface, which depends on the density contrast between the ambient plume and the interface, normalized by the density boundary layer thickness:
\begin{equation}
\dot{m} =  \gamma_\rho \, \theta \, \frac{\rho_\mathrm{plume}-\rho_i}{\delta_\rho} \,,
\label{eq:melt_param}
\end{equation}
where $\gamma_\rho$ is a transfer coefficient, $\theta$ is the ice interface slope, $\rho_\mathrm{plume}$ is the density of the turbulent plume adjacent to the interface, $\rho_i$ is the interfacial density, and $\delta_\rho$ is the density boundary layer thickness.

The interfacial density $\rho_i$ is determined diagnostically from the three-equation melt parameterization, which enforces thermodynamic equilibrium at the ice-ocean boundary. The interface temperature must satisfy the pressure-dependent freezing point relation $T_i = T_f(S_i)$, while heat and salt fluxes balance melting and mixing processes.

The density of the turbulent boundary-layer plume increases with $z$ as a result of ambient entrainment and vertical transport of meltwater. Near the wall ($x \lesssim \delta_u$), the plume is dominated by newly produced meltwater and has a density close to $\rho_i$. Farther from the interface, turbulent mixing progressively incorporates warmer, saltier ambient fluid, causing the plume density $\rho_{\mathrm{plume}}$ to approach the far-field value $\rho_\infty = \rho(T_\infty, S_\infty)$. This introduces a subtle feedback: as entrainment alters the plume density, the local buoyancy relative to the surrounding fluid changes, thereby modulating the convective intensity. In principle, $\rho_{\mathrm{plume}}$ can be specified using wall-bounded plume theory \citep{Zhao23GRL}. However, for the small vertical scales considered here, the plume density does not increase substantially and a first-order approximation $\rho_{\mathrm{plume}} \approx \rho_\infty$ is adequate, since the density gradients within the diffusive boundary layer are much larger by comparison.

\subsection{Density Boundary Layer Thickness}

The primary insight from our parameter sensitivity experiments is that the density boundary layer thickness $\delta_\rho$ adjusts dynamically to balance competing eddy fluxes from competing processes: buoyancy-driven convection, externally-imposed shear, and plume-generated shear. We propose that the inverse boundary layer thickness (equivalently, the effective transfer velocity) is determined by the maximum of these three mechanisms:
\begin{equation}
\frac{\gamma_\rho}{\delta_\rho}  = \Gamma_S \, \, \mathrm{max} \left\{ \underbrace{ c_S \theta (\rho_\mathrm{plume}-\rho_i)^{1/3} }_\text{convection}, \, \mathrm{max}\left(\underbrace{ \sqrt{C_d} \, v_\infty}_\text{external shear}, \, \underbrace{u_*^\mathrm{plume}}_\text{plume shear} \right) \right\},
\label{eq:delta_param}
\end{equation}
where $\Gamma_S$ is a turbulent transfer coefficient for salinity (approximately $10^{-3}$--$10^{-2}$ based on our simulations), $c_S$ is a convective scaling constant, $C_d$ is the effective drag coefficient, $v_\infty$ is the external velocity magnitude, and $u_*^\mathrm{plume}$ characterizes friction velocity generated by the buoyant plume itself.

\textit{Convective scaling.} The first term in equation~\eqref{eq:delta_param} represents purely buoyancy-driven boundary layer transport. In the absence of external or plume-driven shear, the boundary layer thickness adjusts such that convective velocities $w \sim [g(\Delta\rho/\rho_0)\delta_\rho]^{1/3}$ balance molecular diffusion. This yields $\delta_\rho^{-1} \sim (\Delta\rho)^{1/3}$, consistent with classical turbulent natural convection scaling \citep{Versteegh1998}, and has been analyzed theoretically for thermal convection in \citet{Grossmann11}. The temperature, salinity, slope angle, and stratification sensitivity experiments directly validate this scaling: for all of these cases as $(\rho_\infty - \rho_i)$ increases, the leading order effect is that the salinity boundary layer thickness decreases, which leads to a melt rate increase. The stratification experiments (N1--N3) exhibit the most modest influence on $\delta_\rho$ near the interface.

\textit{External shear scaling.} The second term captures the thinning of the boundary layer due to externally-imposed currents. Wall-parallel flow at velocity $v_\infty$ generates turbulent shear with friction velocity $u_* \sim \sqrt{C_d} v_\infty$, which enhances mixing and reduces $\delta_\rho$. The external velocity experiments (V1--V5) demonstrate this mechanism: $\delta_S$ decreases systematically from $0.45$ mm at $v_\infty = 0$ to $0.22$ mm at $v_\infty = 25$ cm~s$^{-1}$, approximately consistent with $\delta_\rho^{-1} \sim \sqrt{C_d} v_\infty$. The corresponding enhancement in melt rate illustrates the effect of ambient ocean currents on ice-ocean heat transfer. The transition from convection-dominated to shear-influenced regime occurs between $0 < v_\infty < 5$ cm/s.

\textit{Plume-driven shear scaling.} The third term represents shear generated by the along-interface flow within the buoyant melt plume. Although this mechanism has not been explicitly tested here, we hypothesize that turbulent kinetic energy (TKE) production associated with the outer melt plume (outside the diffusive boundary layer or driven by a discharge plume, when present) can be interpreted as an effective external velocity forcing. This has been analyzed theoretically for thermal convection \citep{Grossmann00}. Under this view, plume-driven shear competes with buoyancy-driven convection in a manner analogous to imposed ambient currents. This hypothesis will be examined in a future study.

The maximum function in equation~\eqref{eq:delta_param} reflects the idea that the most efficient mixing mechanism controls the TKE, which sets the boundary layer thickness through the salinity scalar balance. In quiescent conditions with vertical interfaces (reference case), convection dominates and $\delta_\rho \sim (\Delta\rho)^{-1/3}$. With external currents, shear progressively takes over and $\delta_\rho \sim (\sqrt{C_d} U)^{-1}$.

\subsection{Simplified Parameterization and Validation}

For practical application to our numerical experiments, we employ two simplifying approximations. First, we approximate $\rho_\mathrm{plume} - \rho_i \approx \rho_\infty - \rho_i$, where $\rho_\infty = \rho(T_\infty, S_\infty)$ is the far-field ambient density calculated from the equation of state. This approximation is well-justified for our parameter range where: (i) interfacial salinity differs substantially from both plume and ambient values, maintaining a strong density gradient throughout the boundary layer; and (ii) the density contrast driving interfacial heat flux is fundamentally set by the far-field thermal reservoir. The approximation would deteriorate for very gentle interface slopes ($\theta < 15$ degrees) with weak melt rates, where vertical density stratification within the boundary layer becomes comparable to the interface-to-ambient contrast.

Second, we approximate the density boundary layer thickness by the salinity boundary layer thickness: $\delta_\rho \approx \delta_S$. This is well-founded because salinity anomalies dominate density variations within the boundary layer due to the much larger haline contraction coefficient relative to thermal expansion ($\beta \Delta S \gg \alpha \Delta T$ in the near-wall region). The salinity boundary layer is significantly thinner than the thermal boundary layer ($\delta_S \ll \delta_T$, with $\delta_S/\delta_T \sim Le^{-1/3} \approx 0.2$), and the haline contribution to buoyancy exceeds the thermal contribution in the critical near-wall region where density gradients are steepest. This approximation would become inaccurate for very warm ambient temperatures ($T_\infty \gg 10$ deg.\ C) where thermal expansion becomes comparable to haline contraction in magnitude, or in parameter regimes where double-diffusive instabilities fundamentally alter the scalar boundary layer structure.



\begin{figure}
  \centerline{\includegraphics[width=1.1\linewidth,angle=0,trim={0pc, 0pc, 0pc, 0pc}, clip]{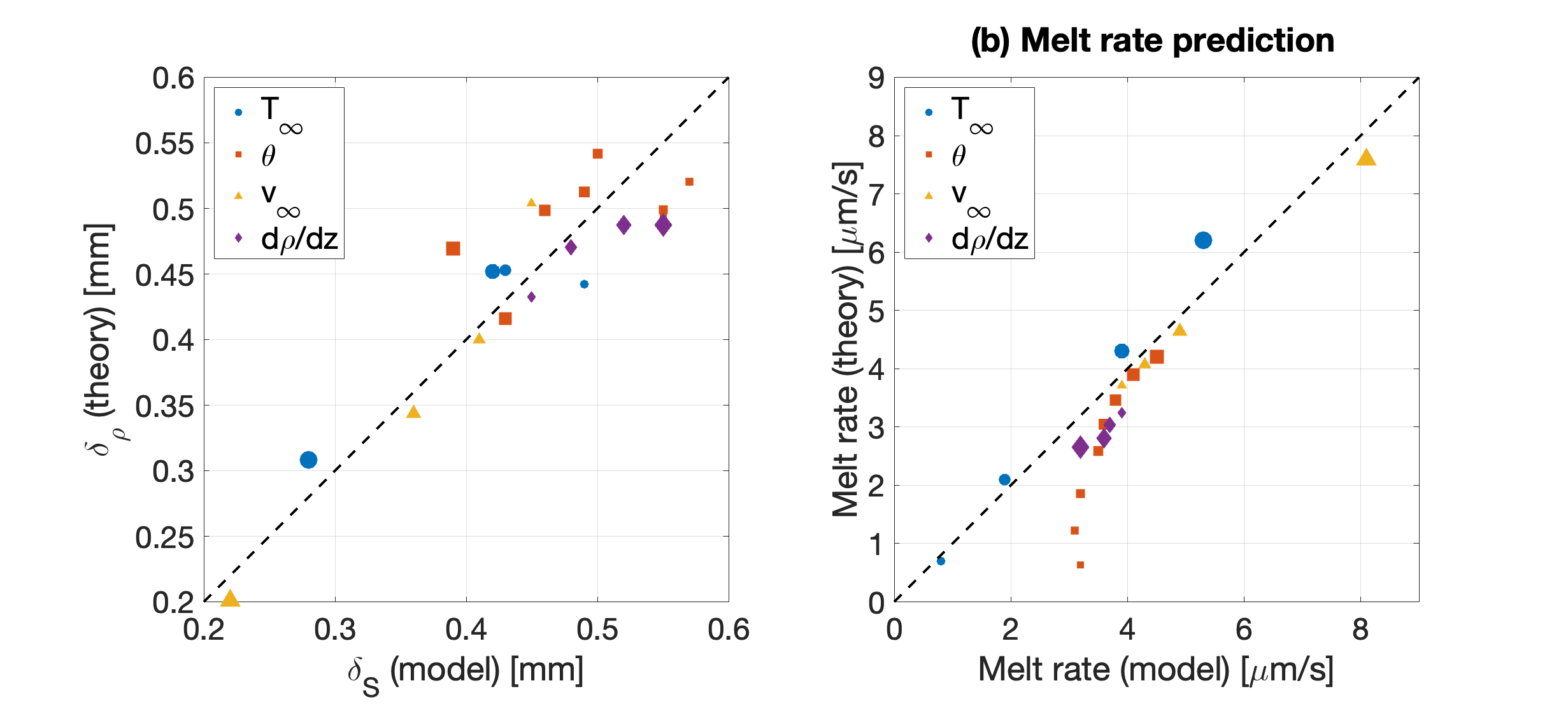}}
  \caption{(a) Model solutal boundary layer thickness $\delta_S$ vs. theoretical density boundary layer thickness $\delta_\rho$. (b) Numerically-diagnosed vs. predicted melt rate ($R^2 = 0.82$). Symbols: $\circ$ $T_\infty$, $\square$ $\theta$, $\triangle$ $v_\infty$, $\diamond$ $d\rho/dz$ with larger symbols corresponding to greater values.}
\label{fig:8}
\end{figure}

Figure~\ref{fig:8} presents a comprehensive validation of the parameterization by comparing numerically-diagnosed melt rates against predictions from equations~\eqref{eq:melt_param}--\eqref{eq:delta_param} across all parameter sensitivity experiments. The parameterization successfully captures the observed trends with coefficient of determination $R^2 = 0.82$ across all cases ($R^2 = 0.87$ excluding shallow slope angles $\theta \leq 45°$), reproducing: (i) the near-proportional increase in $\dot{m}$ with thermal forcing $(\Delta T)^{4/3}$ (T1--T4); (ii) the approximately linear enhancement with external velocity $U$ (V0--V5); (iii) the weak 18\% reduction due to strong stratification (N1--N3); (iv) the dependence on slope angle $\theta$ (S1--S7); and (v) the 10--15\% amplification from moving boundary effects (T1m--T4m). The scatter ($1 - R^2 = 0.18$) arises primarily from the slope angle dependence: at shallow angles ($\theta \leq 45°$) the underlying flow assumptions break down as double-diffusive effects begin to emerge (to be examined in a future study). Despite these limitations, the framework successfully synthesizes diverse parameter sensitivities into a physically-grounded parameterization requiring only far-field properties ($T_\infty$, $S_\infty$, $v_\infty$, $N^2$, $\theta$) and interface thermodynamics, with the key insight that melt rates emerge from competition between buoyancy-driven convection and shear-driven turbulence, where boundary layer thickness adjusts to balance the most efficient mixing mechanism against molecular diffusion.

\subsection{Comparison with Observations}
\label{sec:validation_obs}

The parameterization developed here represents a conceptual advance beyond 
existing frameworks in two respects. First, it unifies the buoyancy-driven 
and shear-driven end members, which were previously treated as distinct regimes 
requiring separate parameterizations, into a single physically-grounded 
expression that transitions smoothly between them for increasing background velocity $v_\infty$. 
Second, it explicitly resolves the role of slope angle $\theta$ and ambient 
stratification $N^2$, which are absent from the widely-used three-equation 
thermodynamic framework \citep[e.g.,][]{Jenkins91}. To illustrate the 
practical significance of these advances, we apply the parameterization to 
two recent fine-scale ice-ocean boundary layer observations.

At LeConte Glacier (Xeitl S\'{i}t'), a near-vertical tidewater glacier in 
southeast Alaska with extensive recent observations \citep{Zhao2024, Nash2024, 
Weiss2025, Ovall2025}, the buoyancy-driven end member ($v_\infty = 0$) of 
our parameterization predicts submarine melt rates of $0.3$--$0.5$~m~day$^{-1}$ 
along near-vertical ice faces, consistent with the lowest end of directly measured melt rates 
reported by \citet{Weiss2025}. As background shear velocity increases, our 
parameterization transitions smoothly toward a shear-dominated regime whose 
scaling converges with the three-equation thermodynamic parameterization 
\citep{Jenkins91} at high $v_\infty$ (2--$3$~m~day$^{-1}$ for $v_\infty$~0.3 m/s), also consistent with the observational 
scaling reported by \citet{Weiss2025}. This convergence validates both the 
high-$v_\infty$ limit of our framework and confirms that the three-equation 
approach remains appropriate when strong background currents dominate over 
buoyancy-driven convection; however it also highlights why the low-$v_\infty$ limit for the three-equation approach with a shear parameterization breaks down in the quiescent limit relevant 
to many tidewater glacier settings.

At crevasses near the grounding line of Thwaites Glacier, we can apply the parameterization using far-field conditions 
representative of the grounding line region \citep[$T_\infty = -0.34\,^\circ$C, 
$S_\infty = 34.3$~psu;][]{Schmidt2023}, assuming a small background shear velocity (which is small, but not as well constrained in this setting). Across slope angles $\theta = 
30$--$90^\circ$, our parameterization predicts melt rates of $17$--$55$~m~yr$^{-1}$, 
with the strong slope dependence arising from the geometric modulation of 
buoyancy-driven convection. These estimates are broadly consistent with 
observational inferences near the grounding line crevasses beneath Thwaites Ice Shelf \citep{Schmidt2023}, and the 
explicit slope sensitivity highlights a potentially important source of spatial 
variability in basal melt that is absent from slope-independent parameterizations. 
The remaining uncertainty at Thwaites is dominated by the unknown $v_\infty$: 
our framework predicts that even modest background velocities ($v_\infty \sim 
5$~cm~s$^{-1}$) could increase melt rates by $30$--$50\%$ above the quiescent 
estimate, motivating targeted observations of near-ice currents in this setting. However, careful melt rate measurements that simultaneously resolve the combined 
convective--shear regime alongside direct observations of all parameters entering 
the parameterization ($T_\infty$, $S_\infty$, $v_\infty$, $N^2$, $\theta$) are 
not commonly available in existing observational datasets, and coordinated 
field campaigns targeting this full parameter set represent an important 
priority for future study.

\section{Discussion and Conclusions}\label{sec:discussion}

This study advances our understanding of turbulent ice-ocean boundary layers by incorporating realistic salinity diffusivities (Schmidt number $Sc \approx 2500$) and systematically exploring the parameter space relevant to Antarctic ice shelf cavities and marine-terminating glaciers in Greenland. Our direct numerical simulations reveal that, for typical glaciological conditions ($0 < T_\infty < 8$ deg.\ C, $25 < S_\infty < 35$ psu), melt and buoyancy fluxes depend on both heat and salt fluxes at the boundary. This occurs because a large Schmidt number creates a thin salinity boundary layer ($\delta_S \approx 0.2$--$0.6$ mm) that generates large density gradients and strong buoyant production of turbulent kinetic energy within millimeters of the ice-ocean interface for ice slopes steeper than approximately 30 degrees from horizontal. 

A key finding is that the salinity boundary layer thickness is the dominant 
constraint on melt rate, set by the balance between convective motions (buoyant 
production of TKE) and outward diffusion of freshwater over most of the parameter 
regime tested (except at larger external shear velocities). This is evidenced by 
our momentum and scalar budget analyses (Section~\ref{sec:budgets}). The 
turbulence closure of Section~\ref{sec:closure} also provides a conceptual bridge 
between this boundary layer dynamics and the melt parameterization: the hierarchy 
$\delta_S \ll \delta_T \leq \delta_\nu$ means that the melt rate is controlled 
primarily by $\delta_S$, which adjusts dynamically to balance the dominant 
mixing mechanism (convective vs. shear-driven) against molecular 
diffusion of salt. Unlike the thermal boundary layer, which may transition from 
convective to shear turbulence regimes from natural convection alone 
\citep{Grossmann11, Yang23, Wells23}, the small molecular diffusivity of salt 
prevents turbulent eddies driven by convection alone from thinning the solutal 
diffusive sublayer (at least on the vertical scales considered), making $\delta_S$ 
the rate-limiting quantity in both the buoyancy-dominated and shear-influenced 
regimes.

Our parameter sensitivity experiments demonstrate that melt rates are sensitive to the retreat of the melting boundary, rate of salt diffusivity, thermal forcing, external velocity, stratification, and interface slope. We synthesize these dependencies into a unified parameterization (Section~\ref{sec:param}) based on the competition between convective and shear-driven processes in balancing molecular diffusivity, which together determines the density boundary layer thickness.
This framework naturally explains the observed trends: thermal forcing increases $(\rho_\infty - \rho_i)$ and thins $\delta_\rho$ through enhanced buoyancy flux and convection; external currents thin $\delta_\rho$ through shear-driven turbulence; and inclined geometries introduce weakened shear and increasing stratification that modifies boundary layer structure.

Importantly, this work reconciles two previously disparate melt parameterizations: those that are depth- or velocity-independent (convectively driven) and those that are velocity-dependent (shear-driven). We show that both regimes naturally coexist in realistic ice--ocean systems, with the dominant mechanism set by the background flow strength.
Under quiescent conditions ($v_\infty \to 0$), buoyancy-driven convection controls the melt, and rates scale with thermal driving as $\dot{m} \sim (\Delta T)^{4/3}$. In contrast, when strong enough ambient currents are present ($v_\infty \gtrsim 5$ cm s$^{-1}$), external shear becomes important and melt rates exhibit the linear velocity dependence $\dot{m} \propto v_\infty$ that is commonly assumed in ocean models.

Some caveats must be considered when applying our parameterizations. First, our results are primarily valid for interface slopes steeper than approximately 15 degrees from horizontal. Gentler slopes require significantly larger computational domains to capture equivalent Rayleigh number regimes. Domain size requirements scale unfavorably with decreasing slope angle, such that reaching equivalent turbulent Rayleigh regimes for near-horizontal interfaces demands computational resources exceeding those employed here. Our drag coefficient and melt rate parameterizations should not be applied to gently-sloping ice shelf regions and requires further study in these scenarios. 
Second, the approximation $\delta_\rho \approx \delta_S$ underlying our parameterization breaks down for very warm ambient temperatures ($T_\infty \gg 10$ degrees C) where thermal expansion becomes comparable to haline contraction. In such regimes, the density boundary layer thickness would be influenced by both thermal and haline dynamics, requiring modification of the transfer coefficient formulations. This limitation may affect parameterizations for Arctic tidewater glaciers or subglacial discharge plumes where ambient temperatures can exceed 10 degrees C.
Third, we have neglected several physical processes that may become important 
in specific contexts: suspended sediment or ice crystals affecting stratification; 
thermohaline interleaving and intrusions at larger scales; tidal and internal wave 
forcing; three-dimensional geometric effects such as basal channels and keels; 
and seasonal evolution of ambient water masses. Each could modify boundary layer 
structure and alter the effective transfer coefficients derived here. Of these, 
the most significant practical gap is the prediction of the background velocity 
$v_\infty$ itself: while our parameterization quantifies how external shear 
modifies melt rates, the magnitude and variability of near-ice currents in 
real glacier and ice shelf settings remains poorly constrained observationally 
and difficult to predict from large-scale models. Since the transition between 
convection-dominated and shear-dominated regimes occurs at modest velocities 
($v_\infty \sim 5$ cm s$^{-1}$), uncertainty in $v_\infty$ likely represents 
the dominant source of error when applying this parameterization in practice.

For climate models, our findings directly address the order-of-magnitude 
discrepancies between observed and parameterized melt rates that arise when 
background currents are weak and convective processes dominate 
\citep{Jackson20, Sutherland19}: by explicitly representing the 
buoyancy-driven end member and its smooth transition to shear-dominated melting, 
the present framework substantially reduces this bias. Implementing our 
physically-based melt rate theory could improve projections of basal melt rates 
under future warming for marine-terminating glaciers in Greenland and portions 
of Antarctic ice shelves, where small melt rate biases translate to large sea 
level uncertainties over long timescales.




\backsection[Funding]{This material is based in part upon work supported by the National Science Foundation Office of Polar Programs Grant OPP-2138790 and OPP-2023674. This work used computational resources supported by NCAR CISL. }

\backsection[Declaration of interests]{ The authors report no conflict of interest.}

\backsection[Data availability statement]{
The Oceananigans model configuration and test case is available at: \textit{https://github.com/zhazorken/iceplume}.
A permanent repository including version model setup, compilation files, and checkpoints will be uploaded to zenodo.org and the Arctic Data Center. }

\backsection[Author ORCIDs]{K. Zhao, https://orcid.org/0000-0002-6260-4315; T. Chor, https://orcid.org/0000-0003-0854-3803; C. McConnochie, https://orcid.org/0000-0001-7105-192X}





\bibliographystyle{jfm}
\bibliography{jfm,references}

\end{document}